\documentclass[11pt]{article}
\pdfoutput=1

\usepackage{jheppub}

\usepackage{mathtools}
\usepackage{amsmath, amssymb, amsthm, amsbsy, bbm}
\usepackage{dsfont}
\usepackage{physics}

\usepackage[percent]{overpic}
\usepackage{xcolor}
\usepackage[utf8]{inputenc}
\usepackage{enumerate}
\usepackage{caption}
\usepackage{subcaption}

\usepackage[normalem]{ulem}
\usepackage{etoolbox}
\usepackage{cancel}
\usepackage{xargs}                      

\newcommand{\calA}{\mathcal{A}}
\newcommand{\calB}{\mathcal{B}}
\newcommand{\calC}{\mathcal{C}}

\newcommand{\calG}{\mathcal{G}}

\newcommand{\calO}{\mathcal{O}}

\renewcommand{\Im}{\mathrm{Im}\,}

\newcommand{\hyp}{{}_2F_1}
\newcommand{\D}{\mathrm{d}}

\newcommand{\myphi}{\varphi}

\newtheorem{conj}{Conjecture}

\newcommand{\Setw}{\mathcal{S}^m_{ETW}}
\newcommand{\Sads}{\mathcal{S}^m_{AdS}}

\newcommand{\bulkM}{\mathcal{M}} 
\newcommand{\braneB}{\mathcal{B}} 

\newcommand{\bO}[1]{\widehat{\calO}_{#1}} 
\newcommand{\cO}[1]{\calO_{#1}} 

\newcommand{\onept}{\calA} 
\newcommand{\boe}[2]{\calB_{#1 #2}} 
\newcommand{\ope}[3]{\calC^{#3}_{#1 #2}} 

\newcommand{\cftx}{x} 
\newcommand{\cfty}{y} 

\newtoggle{editing}
\togglefalse{editing}

\iftoggle{editing}{%
	\setlength{\marginparwidth}{2cm}
	\usepackage[draft]{showlabels}
	\usepackage[colorinlistoftodos,prependcaption,textsize=small]{todonotes}
    
}{%
	\usepackage[disable]{todonotes}
	
}

\newcommandx{\jamie}[2][1=]{\todo[author=Jamie,linecolor=red,backgroundcolor=red!25,bordercolor=red,#1]{#2}}
\newcommandx{\petar}[2][1=]{\todo[author=Petar,linecolor=blue,backgroundcolor=blue!25,bordercolor=blue,#1]{#2}}
\newcommandx{\david}[2][1=]{\todo[author=David,linecolor=green,backgroundcolor=green!25,bordercolor=green,#1]{#2}}
\newcommandx{\wyatt}[2][1=]{\todo[author=Wyatt,linecolor=purple,backgroundcolor=purple!25,bordercolor=purple,#1]{#2}}
\newcommandx{\chris}[2][1=]{\todo[author=Chris,linecolor=orange,backgroundcolor=orange!25,bordercolor=orange,#1]{#2}}
\newcommandx{\moshe}[2][1=]{\todo[author=Moshe,linecolor=violet,backgroundcolor=violet!25,bordercolor=violet,#1]{#2}}

\usepackage{subfiles}

\title{Looking for (and not finding) a bulk brane}
\author{Wyatt Reeves,}
\author{Moshe Rozali,}
\author{Petar Simidzija,}
\author{James Sully,}
\author{Christopher Waddell,}
\author{and David Wakeham}

\affiliation[\,a]{Department of Physics and Astronomy, University of British Columbia,\\
6224 Agricultural Road, Vancouver, B.C.\ V6T 1Z1, Canada.}

\emailAdd{wreeves@phas.ubc.ca, rozali@phas.ubc.ca, psimidzija@gmail.com, jamie.sully@gmail.com, cwaddell@phas.ubc.ca, david.a.wakeham@gmail.com}

\abstract{When does a holographic CFT with a boundary added to it (a BCFT) also have a `good' holographic dual with a localized gravitating end-of-the-world brane? We argue that the answer to this question is almost never. By studying Lorentzian BCFT correlators, we characterize constraints imposed on a BCFT by the existence of a bulk causal structure. We argue that approximate `bulk brane' singularities place restrictive constraints on the spectrum of a BCFT that are not expected to be true generically. We discuss how similar constraints implied by bulk causality might apply in higher-dimensional holographic descriptions of BCFTs involving a degenerating internal space. We suggest (although do not prove) that even these higher-dimensional holographic duals are not generic.}

\begin{document}
\maketitle

\listoftodos[Notes / TODOs / Etc.]
\newpage

\section{Introduction}\label{sec:intro}

Is every consistent theory of quantum gravity a `string theory'? 
There are many ways to attempt to ask this question (or even just to define the terms). 
Since string theory involves non-perturbative higher-dimensional objects or branes, in the context of AdS/CFT one way of asking this question is to study whether, given a holographic conformal field theory (CFT), defect operators in the CFT are described by gravitational branes in the dual bulk. 
That is, does every holographic CFT have a well-behaved spectrum of branes?
As a step in this general direction, in this paper we ask if, given a holographic CFT, every conformal boundary condition is properly described by the bulk gravitational effective field theory, allowing for the addition of a semi-classical `end-of-the-World (ETW)' brane\footnote{In the literature, the term ``end-of-the-world brane" is sometimes also applied to higher-dimensional duals of holographic BCFTs in which the spacetime caps off smoothly over large distances due to shrinking internal space directions; see e.g. \cite{VanRaamsdonk:2021duo}.
In this work, we will use this term to connote a localized, semi-classical gravitating brane which constitutes a boundary for a given spacetime.}.

A closely related---but more concrete---ambition is to sharpen the holographic dictionary for boundary conformal field theories (BCFT). That is, given a holographic CFT, what additional assumptions---if any---must be made for an associated BCFT to be described by the bulk gravitational effective field theory (again allowing for the addition of semi-classical branes)? 
And can we explicitly write the mapping between solutions of the boundary bootstrap and semi-classical bulk+brane actions?

Sharpening the holographic dictionary for BCFTs is timely. 
Recent works \cite{island,eastcoast,westcoast,inforad,Geng_2020,geng2021information,geng2021inconsistency,chen2020quantum,chen2020quantum2,neuenfeld2021dictionary,neuenfeld2021double} have employed a BCFT as a model of a lower-dimensional gravitational system coupled to an auxiliary CFT. 
A BCFT is then a concrete and calculable model for studying Euclidean wormholes and islands. 
In these works, it has been assumed that the BCFT has a good holographic dual with an ETW brane. 
Furthermore, it has been suggested that one might be able to minimally UV-complete coarse-grained gravitational theories by adding ETW branes to the theory \cite{HarlowJafferis,GaoJafferisRecent}. 
But just how realistic or typical are well-behaved ETW branes in a theory of gravity?

A similar program for sharpening the duality between CFTs and bulk gravitational effective field theory was initiated in \cite{Heemskerk:2009pn}. 
There, the conformal bootstrap was used to argue that any CFT such that
\begin{enumerate}[(i)]
    \item simple correlators factorize in a $1/c$ expansion; and 
    \item the spectrum is gapped such that below some large $\Delta_{\mathrm{gap}}$ the only operators are simple light operators and their multi-trace composites
\end{enumerate}
is dual to a bulk theory of semi-classical Einstein gravity. 
A great deal of subsequent work on the holographic bootstrap has strengthened and refined this claim, for example \cite{Penedones:2010ue,Komargodski:2012ek,El-Showk:2011yvt,Fitzpatrick:2012yx,Afkhami-Jeddi:2016ntf,Kulaxizi:2017ixa,Meltzer:2017rtf,Costa:2017twz,Camanho:2014apa,Belin:2019mnx,Kologlu:2019bco,Caron-Huot:2021enk}.

To begin the parallel program for BCFTs, we first note that the holographic CFT bootstrap typically begins with the assumption that the bootstrap can be studied in a $1/c$ expansion about a universal mean field theory solution (MFT) determined by the CFT two-point function. 
For example, a scalar four-point function would have the schematic form
\begin{equation}
  \expval{\phi \phi \phi \phi}   = \sum \expval{\phi \phi}_{Univ}^2 + \order{1/c} \, ,
\end{equation}
where $\expval{\phi \phi}_{Univ}$ is the universal CFT two-point function. 

In contrast to a CFT, the BCFT two-point function is not universal and  kinematically behaves similarly to a CFT four-point function \cite{Cardy:1984bb,McAvity:1995zd,Liendo:2012hy}. 
Moreover, unlike for the holographic CFT, there is no restriction from the BCFT or its gravitational dual that this two-point function should be perturbatively close to a known universal solution like the MFT. 
Thus, before attempting to understand an analogous correspondence between bulk+brane interactions and perturbative solutions to the BCFT bootstrap, we must first understand the leading order, non-perturbative backreaction of the boundary on the bulk gravitational solution. 

\begin{figure}
    \centering
    \includegraphics[width=0.9\textwidth]{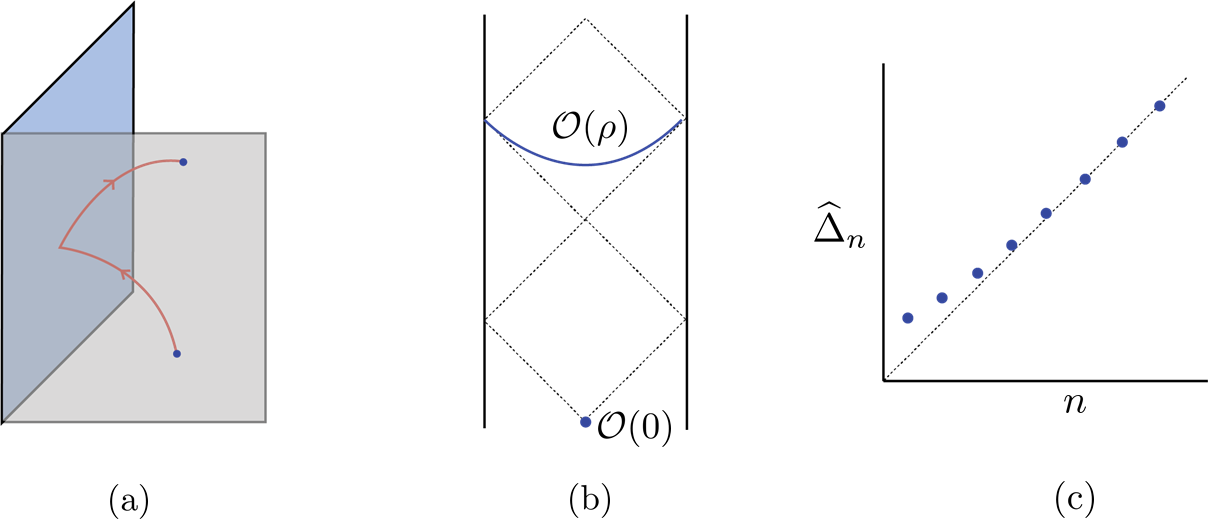}
    \caption{(a) A light ray leaving the boundary and returning to the boundary at a later time (in this example reflecting off an ETW brane); (b) The bulk causal structure then implies new `bulk brane' singularities in the BCFT to the future of a BCFT operator; (c) The bulk brane singularities require a careful alignment of operator dimensions appearing on the boundary of the BCFT.}
    \label{fig:intro-summary}
\end{figure}

To understand what is special about the leading order solution for a BCFT with a simple bulk dual, we will argue that it is useful to rotate to Lorentzian signature, since the Lorentzian BCFT two-point function can probe the \emph{bulk causal structure}. 
When the BCFT has a semi-classical gravitational dual, the bulk causal structure often implies the existence of new approximate singularities in the BCFT\footnote{We don't expect these to be true singularities of the BCFT. Rather, much like the semi-classical singularities predicted by scattering at a bulk point, these will be flattened out at the scale of the gap when the gravitational theory becomes non-local \cite{Maldacena:2015iua}. A similar phenomenon is observed in the two-point function of a holographic CFT at finite temperature; singularities due to null geodesics which wrap the photon sphere of the bulk black hole are resolved by tidal effects in string theory \cite{Dodelson:2020lal}. } (see Figure \ref{fig:intro-summary}). 
Similar singularities for scattering at bulk points in a CFT have been noted before, and their CFT origins were discussed in detail in \cite{Maldacena:2015iua}.\footnote{Lorentzian CFT correlation functions and the singularities from bulk points have been used as a powerful diagnostic of bulk geometry \cite{Engelhardt:2016wgb,Engelhardt:2016vdk,Hernandez-Cuenca:2020ppu}.}

On the BCFT side, these new bulk singularities can only be obtained through the careful alignment of boundary operator dimensions over some large range of dimensions up to a `boundary gap' $\widehat \Delta_{\mathrm{gap}}$. 
The careful alignment of these operators makes such a  bulk causal structure fragile. 
We find no constraints from the CFT being holographic that fix these specific dimensions. 

From the fragility of the bulk causal structure, we suggest that holographic boundary conditions are sparse in the space of all boundary conditions for a holographic CFT. 
On top of the assumptions already necessary for our CFT to be holographic, we must further make a new set of assumptions about the boundary condition itself. 
Namely, we would like to conjecture that a holographic CFT with a boundary condition whose
\begin{enumerate}[(i)]
    \item correlators factorize in an expansion about a (non-universal) free bulk solution; and
    \item boundary operator spectrum is gapped such that below some large $\widehat \Delta_{\mathrm{gap}}$ the only operators are simple light operators and their multi-trace composites;
    \end{enumerate}
is dual to a bulk theory of semi-classical gravity with the possible addition of an ETW brane with a local action. 
It is the first of these conditions that this paper suggests is not generic and must be assumed, although there are subtleties related to this point that appear when we study more complicated top-down constructions of holographic BCFTs. 
The second condition, and the necessity and sufficiency of these two conditions together, will \emph{not} be addressed in this paper. 

This paper is structured as follows. 
We begin with a review of BCFTs and their holographic duals in Section \ref{sec:review}, as well as establishing notation to be used in the rest of the paper. 
A reader familiar with BCFTs can easily skip this section or reference it when the notation we use is not obvious. 
To understand what a holographic BCFT looks like in terms of its spectrum of boundary operators, in Section \ref{sec:simple-model} we examine the simplest possible model: empty AdS cut off by an ETW brane. 
We take the operator spectrum found in our simple model and explain its meaning in Section \ref{sec:lorentz} by studying the bulk causal structure and the Lorentzian continuation of BCFT two-point functions. 
We establish a general correspondence between the boundary operator spectrum and the bulk causal structure in Section \ref{sec:bulk-brane}; this leads to our conjecture about necessary and sufficient conditions for a holographic BCFT. 
We examine our conjecture in top-down models and introduce some necessary caution regarding the strongest version of our claims in Section \ref{sec:beyond-etw-branes}. We conclude with a discussion in Section \ref{sec:discussion}. 

\section{Review of BCFT}\label{sec:review}

Critical phenomena involving a boundary are described by boundary conformal field theories, which involve generalizations of the many familiar concepts and tools of conformal field theories. 
To arrive at a BCFT, one typically introduces a boundary to a known CFT (i.e. we have a finite slab of material). 
One may also introduce additional degrees of freedom living on the boundary, which can be coupled to the CFT degrees of freedom. A complete specification of the theory then involves  imposing boundary conditions for the the bulk degrees of freedom, as well as dynamics for the boundary excitations. 
If this can be done in a manner that maximally preserves conformal invariance, or by flowing to a conformal fixed point, the resulting theory is a BCFT. 
For a given CFT, there may be many different possible choices of conformally-invariant boundary conditions (or conformal fixed points), each of which is described by a different BCFT. 

\begin{figure}
    \centering
    \includegraphics[width=0.4\textwidth]{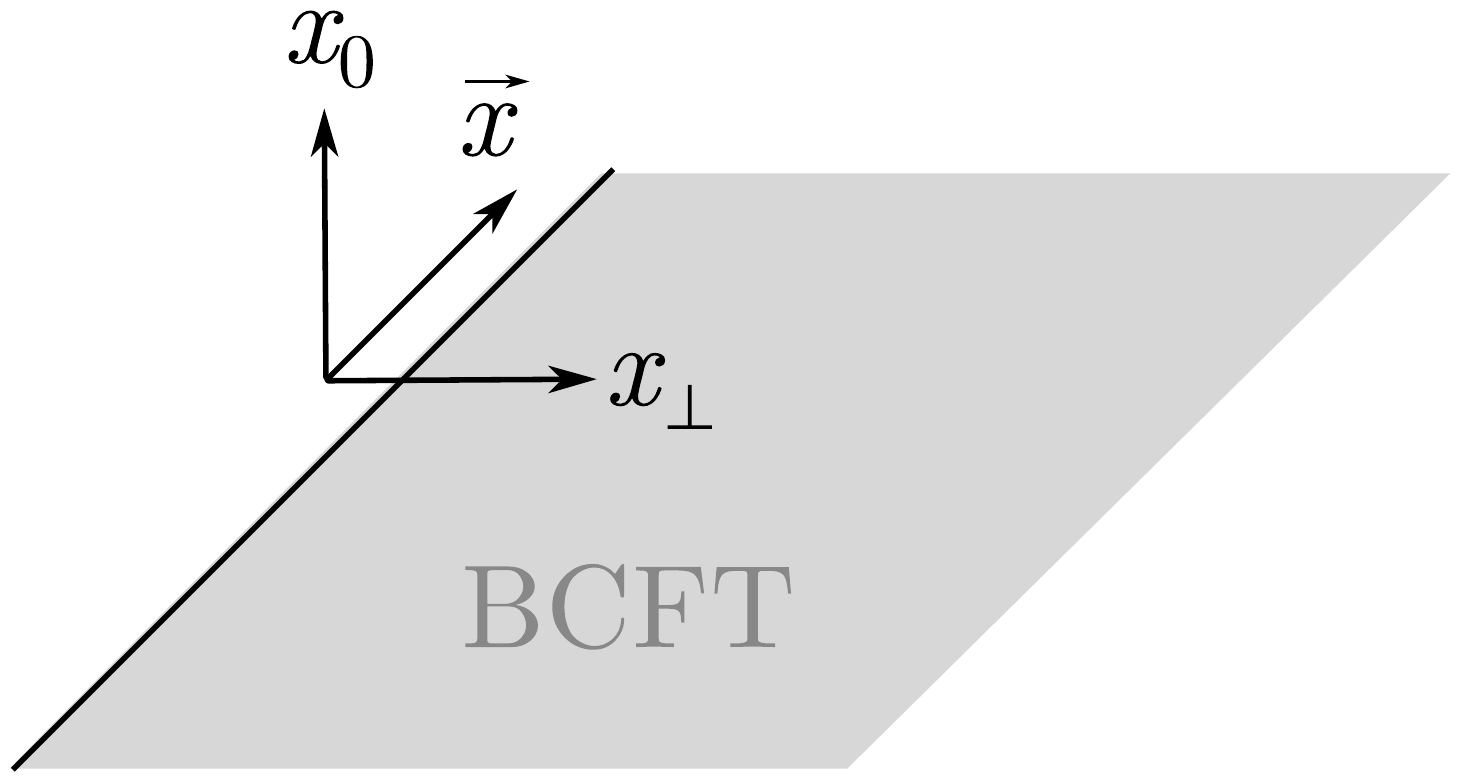}
    \caption{A BCFT on a half-plane $\mathbb{R}^{d-1} \times \mathbb{R}^+$. Here, $x_0$ and $\vec{x}$ are coordinates parallel to the boundary; $x_\perp $ is a coordinates perpendicular to the planar boundary, which sits at $x_\perp=0$.}
    \label{fig:bcft-coordinates}
\end{figure}

\paragraph{Symmetries} 
The most basic tool in studying a BCFT is conformal representation theory: 
the excitations of the theory organize themselves into representations of a reduced conformal symmetry group that is left unbroken by the new boundary. 
When the BCFT lives on the half-plane $\mathbb{R}^{d-1} \times \mathbb{R}_+$ with a planar boundary, the unbroken symmetry is $\mathrm{SO}(d,1) \subset \mathrm{SO}(d+1,1)$, which is the set of transformations that maps the half-plane back to itself. 
We will use coordinates on this space given by $\cftx=(x_0,\vec{x},x_\perp )$, where $x_0,\vec{x}$ are Euclidean coordinates parallel to the boundary and $x_\perp$ is our coordinate orthogonal to the planar boundary. 
We depict these coordinates in Figure \ref{fig:bcft-coordinates}.

\paragraph{CFT Operators and Boundary Operators} 
Because a BCFT only modifies the CFT along the boundary, the spectrum of CFT operators and their algebra remains unchanged. 
Localized on the boundary, however, we have new \emph{boundary operators}, $\hat{\calO}_I$. 
These operators are organized into representations of $\mathrm{SO}(d,1)$, which are partially labeled by a boundary conformal dimension $\hat{\Delta}_I$. 
The boundary conformal dimension is just the usual eigenvalue of the unbroken $d$-dimensional dilation operator (which dilates both along and away from a point on the boundary). 
As is familiar, any such representation has a primary and descendants and we use this structure to organize our description of the BCFT in much the same way as we do for CFTs. 

\paragraph{State-Boundary Operator Map} 
By the usual logic of the state-operator mapping, there is a one-to-one map between boundary operators of the BCFT and states of the theory quantized on a half-sphere ${D}^{d-1}$. 
This follows from the half-plane picture by using an infinite dilation to map back to a point on the boundary. 
Alongside the state-boundary operator map, we still also have the regular CFT state-operator map when we quantize the theory on a sphere $S^{d-1}$ which does not intersect the boundary. 

The state-boundary operator map allows us to write a \emph{boundary operator expansion} (BOE), whereby any CFT operator can be written as a sum over boundary operators
\begin{equation}
    \cO{i}(\cftx) = \frac{\onept_\calO}{(2x_\perp)^\Delta} + \sum_J \frac{\boe{i }{J}}{(2x_\perp)^{\Delta_i - \Delta_J}} \widehat{C}[x_\perp,\partial_{\vec{x}}] \bO{J}(x_0,\vec{x}) \, ,
    \label{eq:boe}
\end{equation}
where the sum over $J$ is over boundary primary operators and the differential operator $\widehat{C}$ which contributes the contributions of descendants is fixed by conformal invariance. 
Likewise, the BCFT inherits the regular OPE from the CFT without a boundary:
\begin{equation}
    \cO{i}(x)\cO{j}(y) = \sum_k \frac{\ope{i}{j}{k}}{|x-y|^{\Delta_i + \Delta_j -\Delta_k}} C[x-y,\partial_y]\cO{k}(y) \, .
    \label{eq:ope}
\end{equation}

\paragraph{Correlation Functions}
Because of the reduced symmetry, BCFT correlation functions involving CFT operators away from the boundary are less constrained than those of a CFT without a boundary. 
A useful `trick' for characterizing the kinematic constraints on a BCFT correlator is to view the correlator as doubled with operator insertions mirrored across the boundary (each copy carrying half the conformal weight of the original operator). 

Following the logic of doubling, one can easily see that a scalar CFT operator has a one-point function that behaves kinematically like a CFT two-point function
\begin{equation}
    \langle \calO(\cftx) \rangle = \frac{A_\calO}{(2x_\perp)^\Delta} \, ,
\end{equation}
where the coefficient $A_\calO$ which determines the size of the vacuum expectation value is a free parameter of the theory, unlike in a CFT, because we choose not to change the normalization of our CFT operators. 

Likewise, the two-point function of scalar operators in a BCFT behaves much like a CFT four-point function and thus no longer fixed by conformal invariance. 
It can be written in terms of an undetermined function of a single conformally-invariant cross-ratio,
\begin{equation}
\langle\calO(\cftx)\calO(\cfty)\rangle = \frac{1}{|4x_\perp y_\perp|^\Delta} \mathcal{G}(\xi)\;,
\label{eq:four-point-fn}
\end{equation}
where the cross-ratio can be defined as
\begin{equation}
    \xi = \frac{(\cftx - \cfty)^2}{4x_\perp y_\perp} = \frac{(x_0 - y_0)^2 + (\vec{x} - \vec{y})^2 + (x_\perp - y_\perp)^2}{4x_\perp y_\perp} \, .
    \label{eq:cross-ratio}
\end{equation}


\subsection{Boundary Bootstrap}\label{sec:rev-bootstrap}

The function $\mathcal{G}(\xi)$ that appears in \eqref{eq:four-point-fn} must decompose into irreducible representations of the conformal symmetry. There are two ways to perform this decomposition. 
We can take the operators near to each other, $\xi \rightarrow 0$, and use the CFT OPE to fuse the two operators into a sum of bulk operators. 
We can then evaluate the sum over local operators in the BCFT. 
The result is an expansion in terms of bulk conformal blocks $g^B$ \cite{Mazac:2018biw,Liendo:2012hy,McAvity:1995zd}: 
\begin{equation}
    \mathrm{Bulk\; Channel:} \quad \quad \mathcal{G}(\xi) = \sum_{i}a_{i} \; g^B_{\Delta_{i}}(\xi) \;,
\end{equation}
where $i$ labels CFT bulk primaries and the coefficients $a_{i}$ are the product of the bulk OPE coefficient and one-point function coefficient of $\calO_i$, 
\begin{equation}
    a_{i} = \ope{}{}{i}\onept_{i} .
\end{equation}
Alternatively, we can take the operators to the boundary, $\xi \rightarrow \infty$, and use the BOE to expand each operator as a sum of boundary operators. We then evaluate the two-point functions of the resulting summed boundary operators, which are fixed by conformal invariance. 
The result is an expansion in terms of boundary blocks $g^b$ \cite{Mazac:2018biw,Liendo:2012hy,McAvity:1995zd}: 
\begin{equation}\label{eq:boundary-block-expansion}
    \mathrm{Boundary\; Channel:} \quad \quad \mathcal{G}(\xi) = \sum_{I}b_{I}\; g^b_{\widehat{\Delta}_{I}}(\xi) \;,
\end{equation}
where $I$ labels boundary primary operators and $b_I$ is the square of their BOE coefficients
\begin{equation}
    b_{I} = {\boe{}{I}}^2 \, .
\end{equation}

\begin{figure}
    \centering
    \includegraphics[scale=0.4]{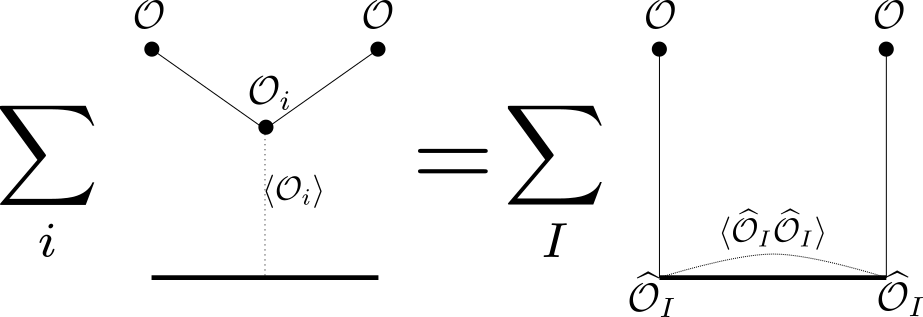}
    \caption{Pictorial representation of \ref{eq:bcft_crossing}. The thick line represents the boundary; thin lines represent fusion of external operators into bulk or boundary operators; dotted lines represent correlators.}
    \label{fig:crossing-symm}
\end{figure}

The equivalence of the expansions in terms of either the boundary conformal blocks or bulk conformal blocks,
\begin{equation}\label{eq:bcft_crossing}
 \sum_{\calO'}a_{i} g^B_{\Delta_{i}}(\xi) = \sum_{I}b_{I} g^b_{\widehat{\Delta}_{I}}(\xi)\;,
\end{equation}
is a BCFT version of \emph{crossing symmetry} and gives bootstrap equations that can be studied with analogous tools as in the CFT case \cite{Liendo:2012hy,Mazac:2018biw}. 
We depict the crossing symmetry visually in Figure \ref{fig:crossing-symm}.

\paragraph{Scalar Blocks}
As shown in \cite{McAvity:1995zd}, the scalar conformal blocks, obtained by solving the Casimir equation for the full and reduced conformal symmetry, are
\begin{align}
    g^B_{\Delta}(\xi) & = \xi^{\Delta/2-\Delta_{\text{ext}}}\hyp\left(\frac{\Delta}{2}, \frac{\Delta}{2}; \Delta - \frac{d}{2} + 1; -\xi\right) \\
    g^b_{\widehat{\Delta}}(\xi) & = \xi^{-\widehat{\Delta}}\hyp\left(\widehat{\Delta}, \widehat{\Delta} - \frac{d}{2}; 2\widehat{\Delta} +2 - d; -\xi^{-1}\right)\;,
    \label{eq:boundary-block}
\end{align}
where $\Delta_{\text{ext}}$ is the dimension of the external operators.

There are branch point singularities in $\mathcal{G}(\xi)$ at $\xi \rightarrow 0, \infty$. As in \cite{Mazac:2018biw}, we take the branch cut to run from $(-\infty, 0)$, so that the Lorentzian continuation $\xi$ lives on the cut plane $\mathbb{C}\backslash(-\infty, 0)$.

\paragraph{Radial Coordinates}
We can also introduce radial coordinates \cite{Pappadopulo:2012jk,Hogervorst:2013sma}, which will simplify some of our expressions:
\begin{equation}
    \xi = \frac{(1- \rho)^2}{4\rho}\;. \label{eq:radial}
\end{equation}
This takes the cut $\xi$-plane to the unit disk $|\rho| < 1$, with $\xi \in (0, \infty)$ mapped to $\rho \in (0, 1)$.
The boundary block is then
\begin{align}
    g_{\widehat{\Delta}}^{b}(\rho) &=\left[\frac{4 \rho}{(1-\rho)^{2}}\right]^{\widehat{\Delta}}{ }_{2} F_{1}\left(\widehat{\Delta}, \widehat{\Delta}-\frac{d}{2}+1 ; 2 \widehat{\Delta}+2-d ; \frac{-4 \rho}{(1-\rho)^{2}}\right) \\
    & = (4 \rho)^{\widehat{\Delta}}{ }_{2} F_{1}\left(\widehat{\Delta}, \frac{d-1}{2} ; \widehat{\Delta}-\frac{d}{2}+\frac{3}{2} ; \rho^{2}\right)\;, \label{eq:rho-bblock}
\end{align}
where on the second line, we used a quadratic transformation.


\subsection{Holographic BCFT}\label{sec:rev-holo}

\subsubsection{Bottom-Up Models}\label{sec:rev-holo-bottom}

In \cite{Fujita:2011fp,Takayanagi2011} (following \cite{Karch:2000ct,Karch:2000gx}) it was proposed that the holographic dual of a BCFT should be a bulk geometry, $\bulkM$, terminated by an ETW brane, $\braneB$, that acts as an additional infrared boundary for the gravitational theory. The new boundary $\braneB$ meets the standard asymptotically AdS boundary at the location of the BCFT boundary (see Figure \ref{fig:simple-model}).
The gravitational sector of the bulk+brane theory is proposed to have an action that now includes a standard Gibbons-Hawking boundary term on the brane
\begin{equation}\label{eq:bulk-grav-action}
S_G=\frac{1}{16 \pi G_{N}} \int_{\bulkM} \sqrt{-g}(R-2 \Lambda)+\frac{1}{8 \pi G_{N}} \int_{\braneB} \sqrt{-h} (K-T)\,,
\end{equation}
where $h$ is the induced metric on the brane, $K$ is the trace of the extrinsic curvature, and $T$ is the tension of the ETW brane. 
One also expects the same bulk AdS matter action as the original CFT without a boundary as well a new matter action living on the ETW brane:
\begin{equation}
S = S_G + \int_M \Sads  + \int_B \Setw  \,.
\end{equation}

The residual $\mathrm{SO}(d,1)$ symmetry of the BCFT fixes the bulk geometry to take the highly-constrained form
\begin{equation}\label{eq:bulk-metric-general}
d s^{2}=d r^{2}+e^{2 A(r)} d s_{A d S_{d}}^{2}\,,
\end{equation}
where a lower-dimensional AdS$_d$ is warped over a radial direction with some warp factor $A(r)$. 
The warp factor is determined by whatever vacuum expectation values are sourced by the ETW matter action $\Setw$, but must asymptotically approach that of empty AdS$_{d+1}$ where $A(r) = \ln\cosh(r)$ as $r\rightarrow - \infty$. 
(We will work in coordinates where the AdS radius $L=1$.)
The ETW brane will sit on some constant radial slice $r=r_0$, fixed by the combination of the tension $T$ and the particular warp factor $A(r)$. 

\subsubsection{Top-Down Models}\label{sec:rev-holo-top}

The bottom-up proposal of \cite{Fujita:2011fp,Takayanagi2011} is known to be too simple to fully describe some explicit top-down constructions of holographic duals that have been derived for `microscopic' BCFTs.  
In these cases, there is a more complicated bulk geometry with a non-trivial internal space. 
The internal space allows the bulk geometry to cap off smoothly in the infrared instead of ending on a brane. 
A few known top-down constructions of holographic BCFT are worth noting here:

\begin{itemize}
    \item In \cite{Chiodaroli:2011fn, Chiodaroli:2012vc}, the authors present general half-BPS solutions of 6-dimensional type 4b supergravity, including those with a single AdS$_{3} \times S^{3}$ asymptotic region, expected to be dual to BCFT$_{2}$. The geometric ansatz for these solutions is a warped product AdS$_{2} \times S^{2}$ fibred over a Riemann surface $\Sigma$. 
The solutions constructed in \cite{Chiodaroli:2011fn} are referred to as the AdS$_{2}$-cap and AdS$_{2}$-funnel; a more general class of solutions is found in \cite{Chiodaroli:2012vc}, involving Riemann surfaces $\Sigma$ with additional boundary components and handles. It is conjectured that these solutions should correspond to supersymmetric configurations of self-dual strings and 3-branes, though this identification and the corresponding class of BCFTs is not well-understood.

\item Half-BPS solutions of 11-dimensional supergravity with a single AdS$_{4} \times S^{7}$ region, corresponding to stacks of semi-infinite M2-branes ending on M5-branes, have been constructed in \cite{Berdichevsky:2013ija, Bachas:2013vza}.

\item In \cite{DHoker:2007zhm, DHoker:2007hhe, Aharony:2011yc}, half-BPS solutions of type IIB supergravity, corresponding to configurations of D3-branes ending on D5-branes and NS5-branes, were constructed. Solutions with a single asymptotic AdS$_{5} \times S^{5}$ region furnish a holographic description of the $U(N)$ $\mathcal{N}=4$ super Yang-Mills theory with $OSp(4|4)$-preserving boundary conditions, classified by Gaiotto and Witten in \cite{Gaiotto:2008sa, Gaiotto:2008ak}.
\end{itemize}

We discuss explicit top-down models in more detail in Section \ref{sec:top-down}.

\subsubsection{Holographic BCFT Dictionary}\label{sec:rev-holo-dict}

Here we review the holographic dictionary for a scalar bulk field in a BCFT. 
We explain how to construct bulk operators in bottom-up models and how to extract their corresponding boundary operator expansion data. 
We follow closely the treatment in \cite{Aharony:2003qf}, although we will use slightly different conventions. 

Consider a bulk scalar field operator $\phi(\vec y, u, r)$ of mass $M$. 
By the standard AdS/CFT dictionary, this field is dual to a CFT operator $\calO_\Delta$ of dimension
\begin{equation}
    \Delta = \frac{1}{2}\left(d+\sqrt{d^2+4M^2}\right) \, .
\end{equation}
At leading order, the bulk field satisfies the free wave equation on the warped background
\begin{equation}\label{eq:Wave_eqn}
    \left(\Box_{\bulkM}-M^2\right)\phi=0 \,.
\end{equation}
We can write a solution of this wave equation in the form 
\begin{equation}
     \sum_n \bar\psi_n(r)\hat\phi_n(\vec y, w),
\end{equation}
where the $\hat\phi_n$ are fields of mass $m_n$ which satisfy the Klein-Gordon equation $\Box_d \hat\phi_n=m_n^2\hat\phi_n$ in AdS${}_d$. 
Substituting the mode expansion into \eqref{eq:Wave_eqn} we find that the radial modes $\bar\psi_n(r)$ must solve
\begin{equation}
    \bar\psi_n''(r)+d A'(r)\bar\psi_n'(r)+e^{-2A(r)}m_n^2\bar\psi_n(r)-M^2\bar\psi_n(r)=0,
\end{equation}
To completely determine the mode expansion, we must also specify a complete set of boundary conditions.
As we are looking for the bulk operator, we require that our solution be normalizable as we approach the AdS boundary
\begin{equation}\label{eq:radial-boundary-limit}
    \bar\psi_n(r) \underset{r\rightarrow  \infty}{=} e^{-  r \Delta}\left(1 + \ldots\right) \, 
\end{equation}
and furthermore choose that the leading term is unit normalized. 
A second boundary condition is specified on the bulk brane, where the specific condition is determined by the terms appearing in the brane action. 
Together, these two boundary conditions determine the correct modes $\psi_n$ and eigenvalues $m_n$, giving the bulk scalar operator
\begin{equation}\label{eq:bulk-operator-exp}
    \phi(\vec y, u, r) = \sum_n \psi_n(r)\hat\phi_n(\vec y, w) \, .
\end{equation}
The last step in writing down \eqref{eq:bulk-operator-exp} is to fix the correct rescaling of the mode functions $\psi_n = c_n \bar\psi_n$. 
The rescaling is determined by enforcing the canonical commutation relations for the bulk field (see Appendix \ref{app:bulk-field-commutators}). 
Note that with this proper normalization in place, the mode functions have the asymptotic form $\psi_n(r) = c_n e^{-  r \Delta} + \ldots $.

Having derived the bulk scalar operator, we can now take the boundary limit to obtain the dual CFT operator. The mode expansion in terms of the AdS$_d$ operators directly gives the dual boundary operator expansion \cite{Aharony:2003qf}.  
However, it is even cleaner to relate the bulk modes to boundary operators by considering the bulk and boundary two-point functions, which we will do in the following. 

\paragraph{Holographic Two-Point Function}

Consider the bulk two-point function in our mode expansion:
\begin{equation}
    \expval{\phi(\mathbf{x}_1) \phi(\mathbf{x}_2)} = \sum_{n,m} \psi_n(r_1) \psi_m(r_2) \expval{\hat \phi_n({x}_1,u_1)\hat \phi_n({x}_2,u_2)} = \sum_{n,m} \psi_n(r_1) \psi_m(r_2) G^{(AdS_d)}_{\Delta_n}({x}_1,u_1;{x}_2,u_2) .
\end{equation}
Using the known form of the AdS$_d$ two-point function (see e.g. \cite{Penedones:2016voo},
\begin{equation}\label{eq:AdS_d-prop}
    G^{(AdS_d)}_{\Delta}(\xi) = \calC_{\Delta,d-1} 2^{-2\Delta} \xi^{-\Delta} \, _2F_1(\Delta,\Delta-d/2+1,2\Delta -d +2,-1/\xi ) 
\end{equation}
where
\begin{equation}
    \calC_{\Delta,d-1} = \frac{\Gamma(\Delta)}{2 \pi^{\frac{d-1}{2}}\Gamma(\Delta-d/2+3/2)} \,,
\end{equation}
one can compute the CFT two-point function in the standard way:
\begin{equation}
    \expval{\calO_1 \calO_2} = \lim_{r_1,r_2\rightarrow \infty} \cosh^{2\Delta}(r) \frac{1}{\calC_{\Delta,d}}\expval{\phi(X_1) \phi(X_2)} .
\end{equation}
Because the AdS$_d$ bulk-to-bulk propagator \eqref{eq:AdS_d-prop} and boundary conformal block \eqref{eq:boundary-block} are identical (up to a constant), we immediately find 
\begin{equation}
     \expval{\calO_1 \calO_2} = \frac{1}{\calC_{\Delta,d}} \sum_n \left(\lim_{r_1,r_2\rightarrow \infty} \cosh^{2\Delta}(r) \psi_n(r_1) \psi_n(r_2)\right) \calC_{\Delta_n,d-1} 2^{-2\Delta} g^b_{\Delta_n}(\xi) ,
\end{equation}
which we can rewrite as 
\begin{equation}\label{eq:bulk-boundary-block-exp}
    \expval{\calO_1 \calO_2} = \frac{1}{\calC_{\Delta,d}} \sum_n c_n^2 \calC_{\Delta_n,d-1} 2^{-2\Delta_n} g^b_{\Delta_n}(\xi) .
\end{equation}

From comparing this expression to \eqref{eq:boundary-block-expansion}, we conclude two things:
\begin{enumerate}
    \item The spectrum of boundary operators appearing in the BOE of $\calO_\Delta$ is given by 
    \begin{equation}\label{eq:boundary-spectrum}
        \left\lbrace \Delta_n =  \frac{1}{2}\left(d-1-\sqrt{(d-1)^2+4m_n^2}\right) \right\rbrace \,.
    \end{equation}
    \item The BOE coefficients are given by
    \begin{equation}\label{eq:boundary-BOE-coeff}
        \calB_n = \frac{1}{2^{\Delta_n}}\sqrt{\frac{\calC_{\Delta_n,d-1}}{\calC_{\Delta,d}}} c_n\;.
    \end{equation} 
\end{enumerate}

\section{Simplest Bulk Model}\label{sec:simple-model}

\begin{figure}
    \centering
    \includegraphics[width=0.5\textwidth]{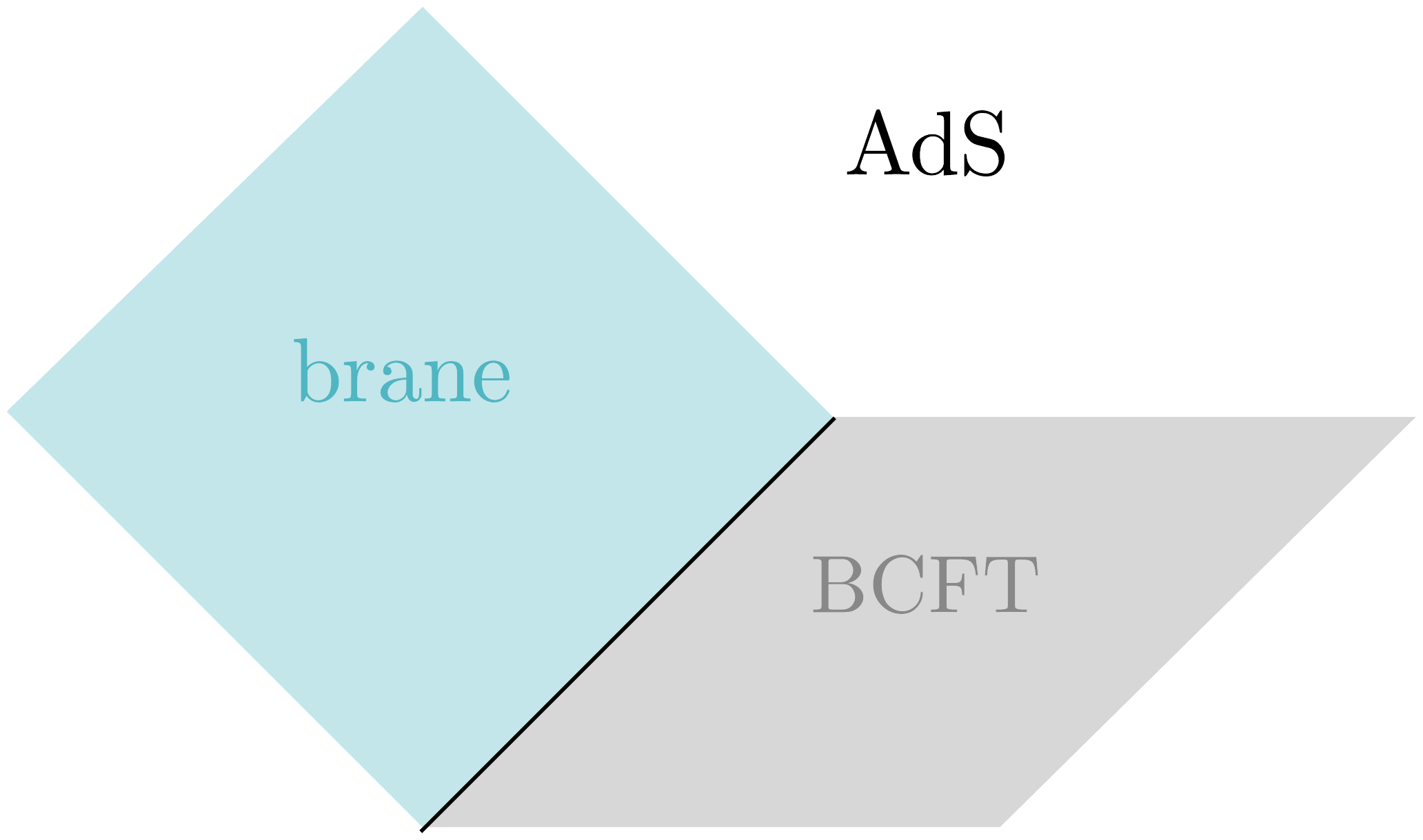}
    \caption{Our simple model in which the bulk is locally AdS$_{d+1}$, but is terminated by an ETW brane. We depict here the AdS$_d$ foliation of AdS$_{d+1}$.}
    \label{fig:simple-model}
\end{figure}

To understand the leading-order free two-point function in a holographic theory, we begin by studying the simplest possible bottom-up model: empty AdS terminated by an ETW brane. 
In our radial coordinates \eqref{eq:bulk-metric-general}, the AdS$_d$ foliation of AdS$_{d+1}$ takes the form 
\begin{equation}
    ds^2 = dr^2+\cosh^2(r)\left(\frac{d\vec y^2+du^2}{u^2}\right).
\end{equation}
The location of the brane is given by some $r=r_0$, determined by the tension. See Figure \ref{fig:simple-model}. 
It will also sometimes be useful to change to an `angular' coordinate using $\tanh(-r)=\cos(\varphi)$ so that
\begin{equation}
    ds^2 = \csc^2\varphi \left[ d\varphi^2+\left(\frac{d\vec y^2+du^2}{u^2}\right)\right] \, , 
    \label{eq:angular-coord}
\end{equation}
simplifying the conformal structure of the metric. 

We consider a free scalar field, $\phi$, whose dual CFT operator has dimension $\Delta$.
We will need to impose boundary conditions at the location of the brane. As a simple choice, we choose Neumann boundary conditions on the field,  $\partial_r\phi(r_0)=0$, although the qualitative features of our results will not depend on this specific choice. 
Using the mode expansion explained in Section \ref{sec:rev-holo-dict},
\begin{equation}
    \phi(\vec y, u, r) = \sum_n c_n \bar\psi_n(r) \hat \phi_n(\vec y, u) \, ,
\end{equation}
we find the two independent radial solutions of the EOM to be
\begin{align}
    \psi_{n(1)}(r) &= \sin^{d/2}(\varphi)P_\nu^\mu (\cos\varphi),\\
    \psi_{n(2)}(r) &= \sin^{d/2}(\varphi)
    \left(\frac{1-\cos\varphi}{1+\cos\varphi}\right)^\frac{\mu}{2}
    {}_2 F_1\left(\nu+1,-\nu,\mu+1,\frac{1-\cos\varphi}{2}\right),
    \label{eq:psi_normalizable}
\end{align}
where we have chosen to use our angular coordinate $\varphi$, while $\nu$ and $\mu$ are
\begin{align}
    \nu &=    \Delta_n-\frac{d}{2},\\
    \mu &=    \Delta-\frac{d}{2}.
\end{align}
The asymptotic behaviour of these solutions as $r\rightarrow \infty$ is
\begin{align}
    \psi_{n(1)} \sim e^{-r(d-\Delta)},\quad \psi_{n(2)} \sim e^{-r\Delta},
\end{align}
which is what we expect from the non-normalizable and normalizable solutions of the wave equation, respectively.

Taking into account our boundary condition on the brane, only the modes $\psi_{n(1)}$ which satisfy $\psi_{n(1)}'(r_0)=0$ are admissible. Since each radial function $\psi_{n(1)}(r)$ is related to a corresponding co-dimension 1 field $\phi_n(\vec y, w)$, with a dimension $\Delta_n$ given by Eq. \eqref{eq:boundary-spectrum}, we expect that the condition $\psi_{n(1)}'(r_0)=0$ will restrict the spectrum of $\Delta_ns$ to take on only a discrete set of values.

In the limit of large $\Delta_n$, we can explicitly solve the equation $\psi_{n(1)}'(r_0)=0$ for $\Delta_n$ to obtain
\begin{align}\label{eq:Delta_n}
    \Delta_n &= \frac{d-1}{2}+
    \frac{\left(n+\frac{1}{2}\left(\Delta-\frac{d}{2}\right)+\frac{1}{4}\right)\pi}{\arccos(\tanh(r_0))}\quad n\in \mathbb{Z}\quad \text{(large $\Delta_n$)}.
\end{align}
In our angular coordinates, where the brane sits at $\varphi_0$, this simplifies to
\begin{align}\label{eq:Delta_n-angular}
    \Delta_n &= \frac{d-1}{2}+
    \frac{\left(n+\frac{1}{2}\left(\Delta-\frac{d}{2}\right)+\frac{1}{4}\right)\pi}{\varphi_0}\quad n\in \mathbb{Z} \, .
\end{align}
Note that, in $d=2$, the position $r_0$ of the brane is understood to be related to the defect entropy $\log g$ of the CFT by~\cite{Takayanagi2011}
\begin{equation}
    \log g = \frac{r_0}{4G},
\end{equation}
which gives
\begin{equation}
    \Delta_n^{d=2} = \frac{1}{2}+
    \frac{\left(n+\frac{\Delta}{2}-\frac{1}{4}\right)\pi}{\arccos(\tanh(2G\log g))}\quad n\in \mathbb{Z}. \quad 
\end{equation}

It remains to calculate the scaling coefficients $c_n$ appearing in \eqref{eq:bulk-operator-exp} by enforcing the equal time canonical commutation relations. Following Appendix \ref{app:bulk-field-commutators}, we compute $c_n$ from the mode normalization
\begin{equation}
    \int_{r_0}^\infty dr \cosh^{d-2}(r)\psi_n(r)\psi_m(r) = \frac{1}{c_n^2} \delta_{nm} \, .
\end{equation}
Using the explicit expressions for $\psi_n(r)$ in Eq. \eqref{eq:psi_normalizable}, we can solve for $c_n$ in the asymptotic limit of large $n$, which is the same limit in which we evaluated the scaling dimensions $\Delta_n$ of the operators $\phi_n$. 
This gives the expression 
\begin{equation}
    c_n
    \approx
    \frac{\pi^{1/4}\sqrt{(1+\mu)}}{2}
    \sqrt{\frac{\Gamma(\mu+1/2)}{\Gamma(\mu+2)\Gamma(2\mu+1)}}
    \left(\frac{2}{\arccos(\tanh r_0)}\right)^{\mu+1}
    \left( 
    \frac{\pi}{4}(4n+2\mu+1)
    \right)^{\mu+1/2},
    \label{eq:c_n}
\end{equation}
with $\mu=\Delta-d/2$, as before.

From Eq. \eqref{eq:boundary-BOE-coeff}, we can then compute the the BOE coefficients by plugging in the above $c_n$ into the expression
    \begin{equation}\label{eq:boundary-BOE-coeff-simple}
        \calB_n = \frac{1}{2^{\Delta_n}}\sqrt{\frac{\calC_{\Delta_n,d-1}}{\calC_{\Delta,d}}} c_n \, .
    \end{equation}
In the large $n$ limit, we can write this as
\begin{equation}
\calB_n = 2^{-\frac{n}{\varphi_0}} n^{\Delta-\frac{d+1}{4}} B + \ldots
\label{eq:boe-simple-model-large-n}
\end{equation}
with $B$ a constant independent of $n$:
\begin{equation}
    B = e^{\frac{3}{4}-\frac{d}{4}} \pi ^{-\frac{d}{4}+\Delta +\frac{1}{2}}
   2^{ \frac{\pi  (d-2 \Delta -1)}{4\varphi }-\frac{d-1}{2}} \varphi^{\frac{d-1}{4}-\Delta
   } \left[ \Gamma(\Delta) \Gamma(-{d}/{2}+\Delta +1)\right]^{-1/2}
\end{equation}

We conclude from this simple model that the information about the bulk geometry, namely---given our restricted assumptions---the location of ETW brane at $\varphi_0$, appears in two places:
\begin{enumerate}
    \item The asymptotic spacing of boundary operator dimensions $\gamma = \lim_{n\rightarrow \infty}\Delta_{n+1} - \Delta_n = \frac{\pi}{\varphi_0}$.
    \item The asymptotic growth of the BOE coefficients $\boe{}{n}\sim \exp\left( \frac{n}{\varphi_0} \ln 2\right)$.
\end{enumerate}
What is not yet clear is why the information about the brane is encoded in this particular way and how it generalizes to a lesson about all BCFTs with good bulk duals. 
To make this next step, we must turn to the Lorentzian structure of two-point functions in a (holographic) BCFT. 

\section{Lorentzian BCFT Singularities}\label{sec:lorentz}
In this section, we will consider the singularities associated with a scalar two-point function in a Lorentzian BCFT. We start by discussing the field theory setup and the expected structure of kinematic singularities.
For BCFTs with a simple holographic dual, we consider the apparent singularities that arise from the bulk causal structure. 
In particular, we consider the bulk null rays that are reflected off the brane, and compute the cross-ratio of the return locus for these rays.

\subsection{BCFT Singularities}\label{sec:sing-bcft}

In Euclidean signature, a CFT correlator has singularities whenever two operators become coincident (and is analytic otherwise). 
Similarly, a Euclidean BCFT correlation function will have singularities only when operators approach each other, or when they approach the boundary. 
(We can think of this as an operator approaching their mirrored double across the boundary.)

In terms of a scalar BCFT two-point function, and our cross ratio $\xi$ defined in \eqref{eq:cross-ratio}, the singularity when the two operators approach each other corresponds to the limit $\xi\rightarrow 0$. 
In this limit, the correlator will diverge like
\begin{equation}
\expval{\calO(\cftx)\calO(\cfty)} = \frac{1}{\abs{\cftx - \cfty}^{2\Delta} }+ \ldots 
\, 
\end{equation}
or, correspondingly, $\calG(\xi) \sim \xi^{-\Delta}$.
When the operators approach the boundary, in the limit $\xi \rightarrow \infty$, the correlator diverges like
\begin{equation}
\expval{\calO(\cftx)\calO(\cfty)} = \frac{\mathcal{A}^2}{\abs{4 x_\perp y_\perp}^{\Delta}} + \ldots 
\, 
\end{equation}
or, correspondingly, $\calG(\xi) \sim \mathcal{A}^2$. 
Unlike the CFT case, there is no third Euclidean singularity, which could be thought to correspond to the operator $\calO(\cfty)$ approaching the mirror of $\calO(\cftx)$.

\begin{figure}
    \centering
    \includegraphics[width=0.5\textwidth]{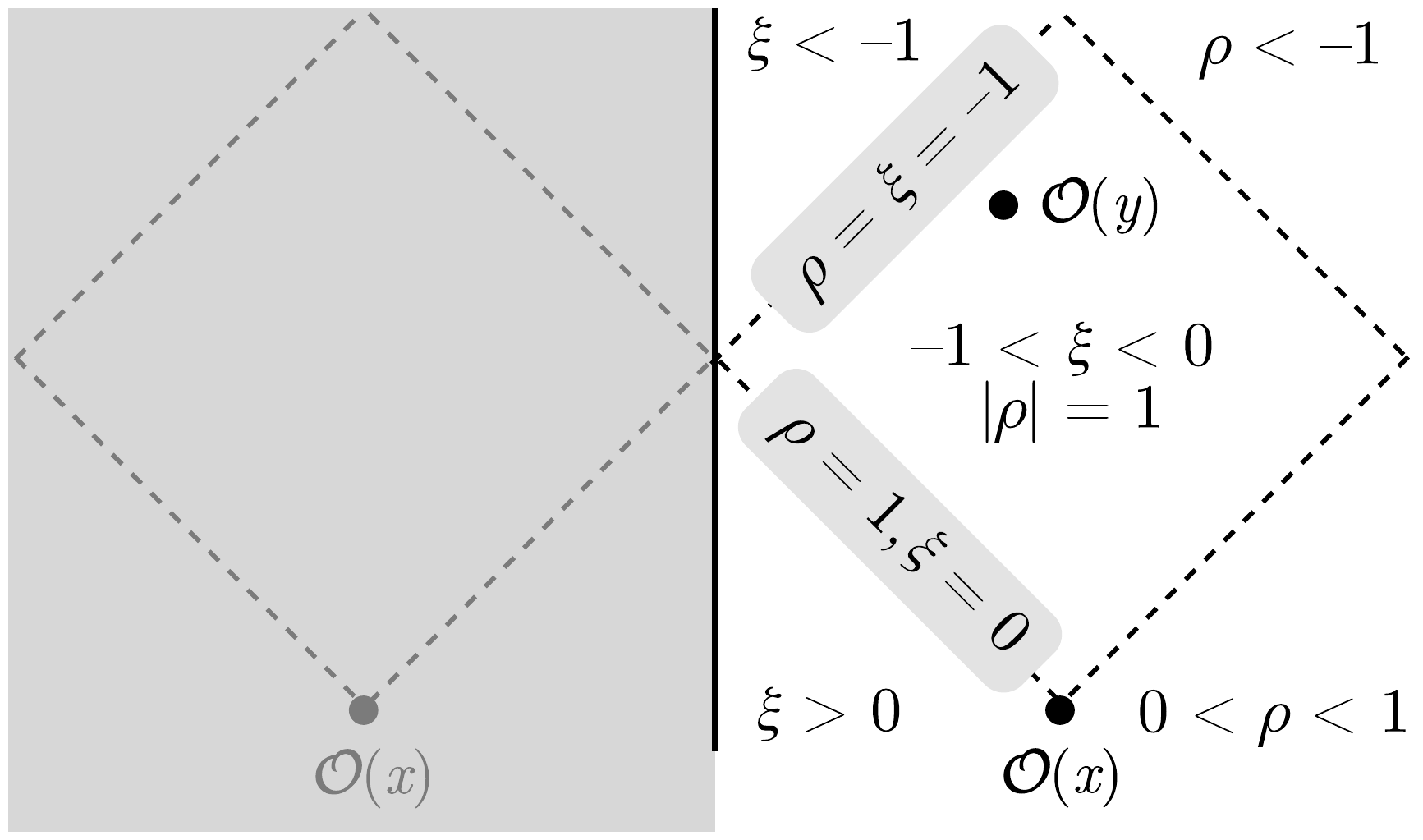}
    \caption{We depict various regions of the Lorentzian interval for a BCFT in terms of various cross-ratios. Importantly we note that the causal diamond bounded by the lightcone of the operator $\mathcal{O}(x)$ and its reflection off the boundary is described by the radial cross-ratio $\rho$ living on the unit circle. It interpolates between the initial lightcone at $\rho = e^{i 0}$ and the reflected ligthcone at $\rho = e^{i \pi}$. }
    \label{fig:lorentzian-causal-structure}
\end{figure}

In Lorentzian signature, we similarly expect a singularity when $\calO(\cfty)$ approaches the lightcone of $\calO(\cftx)$ at the cross-ratio $\xi = 0 $. 
We can also continue the Lorentzian two-point function around the branch point at $\xi=0$ to the timelike region $\xi<0$. 
Here there is another possible singularity where the $\calO(\cfty)$ approaches the reflection of the lightcone of $\calO(\cftx)$ off the boundary at $\xi = -1$. 
This is known as the \emph{Regge Limit} of the BCFT \cite{Mazac:2018biw} and it has been shown that the BCFT diverges here at worst as $\calG(\xi) \sim (\xi+1)^{-\Delta}$. 
This is exactly the singularity one would expect from approaching the lightcone of the `mirror' of $\calO(\cftx)$. We depict the Lorentzian causal structure and the corresponding cross-ratios in Figure \ref{fig:lorentzian-causal-structure}. 

When we change to radial coordinates, placing $\calO(\cfty)$ in the timelike region to the future of $\calO(\cftx)$, but before the reflected lightcone, corresponds to $\rho = e^{i\varphi}$ for $\varphi \in [0,\pi]$. 
At one end $\rho=1$ ($\varphi=0$) is the lightcone of $\calO(\cftx)$ at $\xi=0$ and at the other end $\rho=-1$ ($\varphi=\pi$) is the reflected lightcone of $\calO(\cftx)$ at $\xi=-1$.
We also indicate the $\rho$-regions in Figure \ref{fig:lorentzian-causal-structure}. 


It has been argued that a CFT correlation function should only have singularities at points corresponding to Landau diagrams \cite{Maldacena:2015iua} where null particles interact at local vertices. 
By the same logic, we expect the only singularities of a BCFT two-point function to be that on the lightcone and its reflection. 
We do not not attempt to prove this statement in general, but we can follow \cite{Maldacena:2015iua}, and show that it holds in a 2D BCFT.

\paragraph{BCFT Singularities in 2D}\label{sec:2d}

\begin{figure}
    \centering
    \includegraphics[trim={0 0 15cm 0},clip,width=0.4\textwidth]{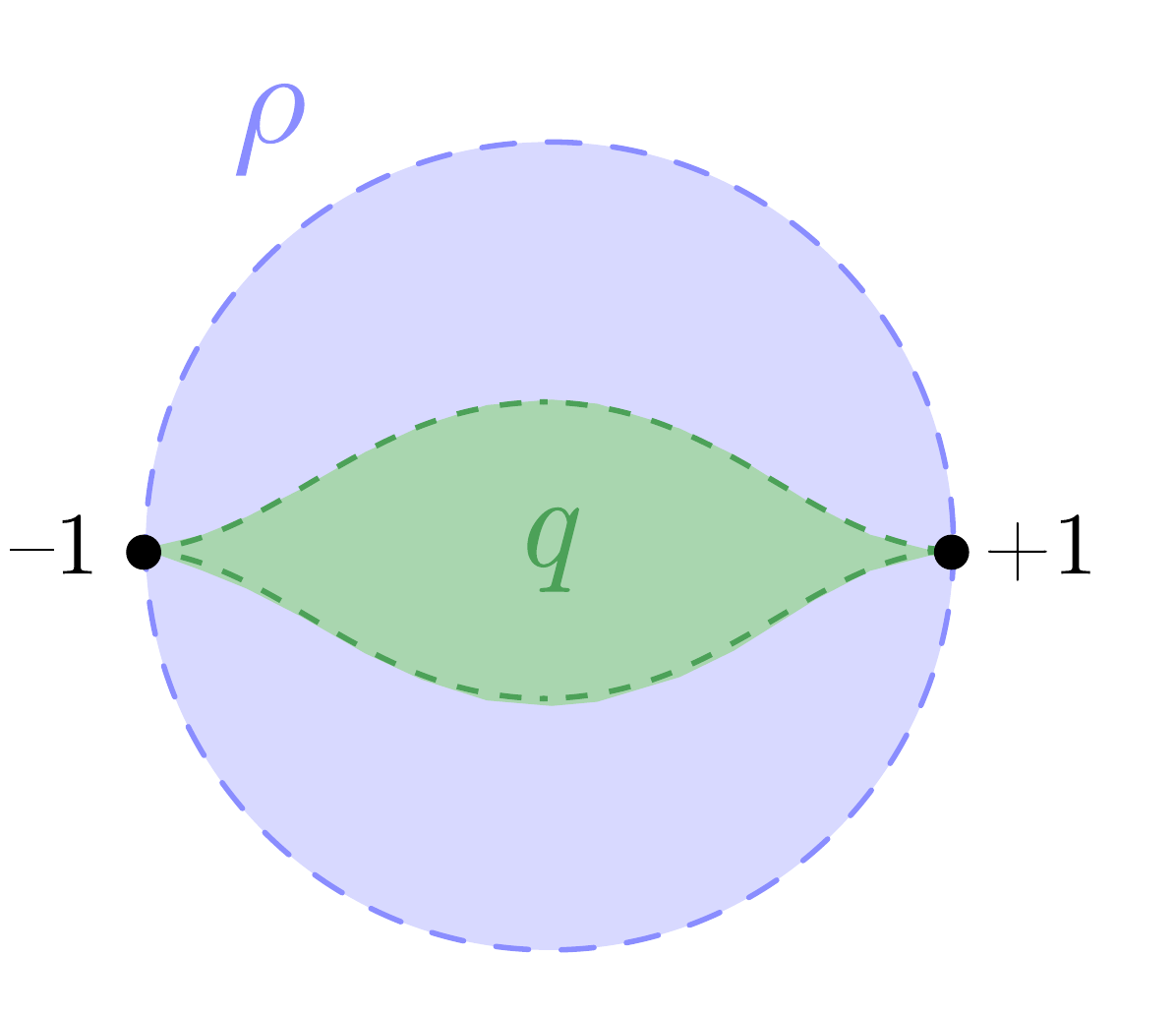}
    \caption{Relation between the radial $\rho$ variable and the new coordinate $q$.}
    \label{fig:q-region}
\end{figure}

In two dimensions one can perform a conformal transformation to map the unit $\rho$ disk into the interior of the unit disk in a new coordinate $q$, hitting the boundary only at $q =  \pm 1$. We depict the region in Figure \ref{fig:q-region}. 
One can then show that the two-point function converges everywhere in the interior of the unit $q$ disk; the convergent region includes the region where the two-operators are timelike separated, except the point $q = \pm 1$ corresponding to the expected BCFT lightcone. We give a more complete derivation of this result in Appendix \ref{app:2d-sing}.

\subsection{Bulk Singularities}\label{sec:sing-bulk}

The location of the BCFT singularities we have just listed are universal and kinematic, in the sense they can be read off the behaviour of individual conformal blocks without consulting the spectrum. 
But for a BCFT with a simple holographic dual, a new type of singularity can emerge: an insertion now generates a lightcone in the \emph{gravitational bulk} as well as the boundary. 
Bulk light-rays can head into the infrared gravitational geometry and return some time later to the boundary, indicating new singularities in the BCFT. 
When the bulk geometry is `shallow' (for example, when the geometry ends on a brane with a large negative tension), these singularities may even occur before the boundary light ray has returned. 

To illustrate this behaviour, we begin by examining our simple toy model where empty AdS is terminated by an ETW brane. 
We can re-write our angular metric \eqref{eq:angular-coord} in the form 
\begin{equation}
    \D s_\text{Euc}^2 = \frac{1}{\sin^2\theta \sin^2\varphi} \left(\D \tau^2 + \cos^2\theta \D  \Omega_{d-2}^2 +\D \theta^2 +  \sin^2\theta\, \D \myphi^2\right)\; , 
    \label{eq:spherical-coords}
\end{equation}
by turning the AdS$_d$-radial coordinate on the slices into a second angular coordinate $\theta$.\footnote{By radial coordinate, we mean the global AdS$_d$-radial coordinate on the slices. These global AdS coordinates can be obtained simply by switching to polar coordinates on the slice, with an origin on the boundary.}  
The angular radial coordinate $\theta$ on the slices takes values $ \theta \in [0, \pi/2]$ with $0$ being the boundary of AdS. 
Recall that the other coordinate $\phi$ takes values in the range $ \varphi \in [0,\varphi_b]$ and is found from the coordinate change $\cos\varphi = \tanh (-r)$. 
$\D\Omega_{d-2}^2$ is the line element on the $\mathbb{S}^{d-2}$ that parametrizes the rest of the AdS$_d$ slice. 
Ignoring the conformal factor, we can see that the angular coordinates $(\theta,\varphi)$ together form part of an $\mathbb{S}^{2}$. 
By continuing to Lorentzian time, we arrive at the metric
\begin{equation}
    \D s_\text{Lor}^2 = \frac{1}{\sin^2\theta \sin^2\varphi} \left(- \D t^2 + \cos^2\theta \D \Omega_{d-2}^2 +\D \theta^2 +  \sin^2\theta\, \D \myphi^2\right)
    \; . 
    \label{eq:spherical-coords-lorentz}
\end{equation}
We will perform our bulk causal calculations in these coordinates.

To begin, we restrict ourselves to consider null rays travelling on the 2-sphere at a fixed position on the $\mathbb{S}^{d-2}$ in \eqref{eq:spherical-coords-lorentz}.
This is a straightforward affair.
Consider a null ray $x^\mu(\lambda) = (t(\lambda), \theta(\lambda), \myphi(\lambda)))$, with affine parameter $\lambda$.
The conformal factor drops out, leaving a simple null geodesic equation
\begin{equation}
    -\dot{t}^2 + \dot{\theta}^2 + \sin^2\theta\, \dot{\myphi}^2 = 0\;.
\end{equation}
We are free to take $\dot{t} = 1$, so that affine time elapsed simply measures distance along the sphere, and the calculation of the return locus reduces to a problem of spherical trigonometry.
Without loss of generality, we take the initial insertion to lie at $x^\mu = (0, \theta_0, 0)$. The null ray will head off into the bulk with some initial direction $\dot{\theta}_0 = \dot{\theta}(0)$, bounce off the brane at $\myphi = \myphi_b$, and return to the boundary $\myphi = 0$ at some angle $\theta_1$ and time $\Delta t = d$ measured by the distance travelled.

\begin{figure}
    \centering
    \includegraphics[width=0.7\textwidth]{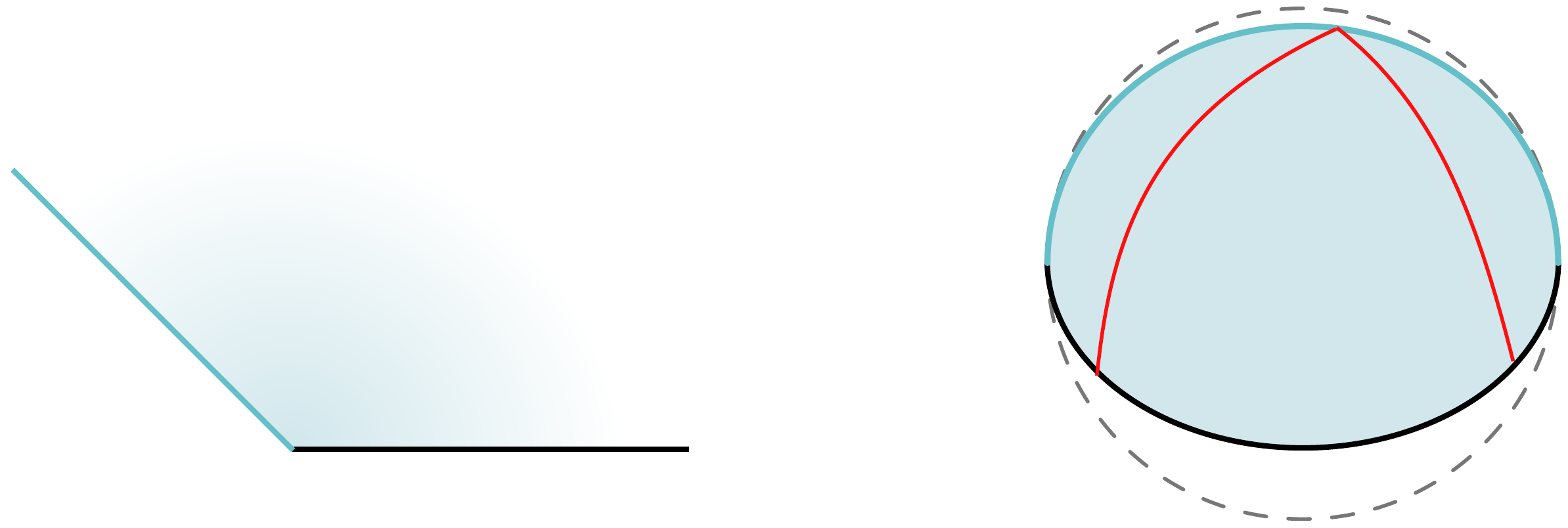}
    \caption{(a) A 2D spatial-slice of AdS$_3$ cutoff by an ETW brane. (b) The same spatial slice conformally mapped to part of the two-sphere. The path of a null geodesic is marked in red.}
    \label{fig:example-null-geodesic}
\end{figure}

To simplify the kinematics further, we can \emph{double} the width of the wedge to $2\myphi_b$. There is now no need to consider the reflection off the brane, since the light ray sails smoothly through the mirror and arrives at the reflected boundary.
It follows immediately from spherical trigonometry\footnote{Specifically, the spherical law of cosines.} that the initial position $\theta_0$, direction $\dot{\theta}_0$, return angle $\theta_1$ and elapsed time $\Delta t = d$ are related by
\begin{equation}
    \cos d 
    = \cos\theta_0\cos\theta_1 + \sin\theta_0 \sin\theta_1 \cos \myphi_b\;. \label{eq:return-time}
\end{equation}
We show the spatial path of one of these null geodesics in Figure \ref{fig:example-null-geodesic}.

\begin{figure}
    \centering
    \includegraphics[width = 0.25\textwidth]{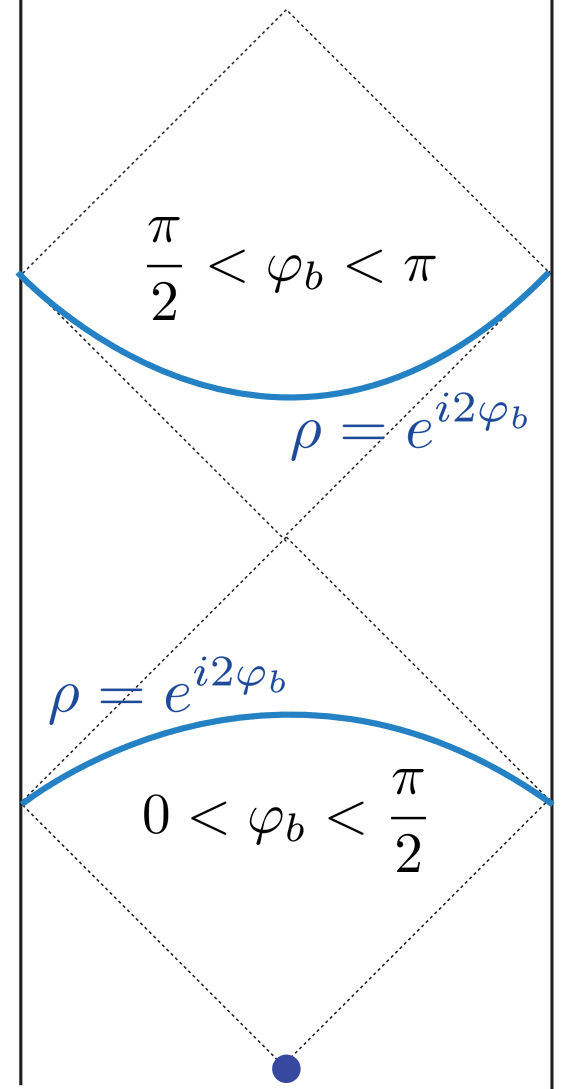}
    \caption{An illustration of two example return loci for branes of different tension/causal depth. When the brane tension is positive, $\varphi_b$ is greater than $\pi/2$ and null geodesics return to the boundary along a curve in the upper causal diamond.  When the brane tension is negative, $\varphi_b$ is less than $\pi/2$ and null geodesics return to the boundary along a curve in the lower causal diamond.}
    \label{fig:return-locus}
\end{figure}

To compute the cross-ratio, $\xi$, for this locus, note that the flat BCFT coordinates are related to our polar coordinates $x_1 = e^{it}\cos\theta$ and $x_\perp = e^{it}\sin\theta$. Plugging in (\ref{eq:return-time}), the analytically continued cross-ratio is
\begin{align}
    \xi 
    & = \frac{(e^{i \Delta t}\cos{\theta_1}-\cos{\theta_0})^2 + (e^{i \Delta t}\sin{\theta_1} - \sin{\theta_0})^2}{4 e^{i \Delta t}\sin{\theta_1}\sin{\theta_0}}
    = - \sin^2\varphi_b \;. \label{eq:locus}
\end{align}
This is pleasingly simple. In terms of our 
radial cross-ratio (\ref{eq:radial}), it is even simpler:
\begin{equation}
    \rho = e^{i 2 \varphi_b}\;.
    \label{eq:return-radial-coord}
\end{equation}
We show the return locus for varying $\varphi_b$ in Figure \ref{fig:return-locus}.

We conclude that the bulk causal structure of our simple ETW brane model predicts a singularity in the BCFT at the cross-ratio \eqref{eq:return-radial-coord}. 
This occurs away from the expected BCFT singularities at $\rho=1,0,-1$. 

\paragraph{General warp factor}\label{sec:return-locus-general-warp}
We can repeat the same argument with minor modifications for a more general warped background plus ETW brane as in \eqref{eq:bulk-metric-general}. 
In this case, one only needs to find the appropriate angular coordinate to put the metric in the form
\begin{equation}
    \D s_\text{Lor}^2 = \frac{1}{\sin^2\theta f(\varphi)} \left(- \D t^2 + \cos^2\theta \D \Omega_{d-2}^2 +\D \theta^2 +  \sin^2\theta\, \D \myphi^2\right)
    \; ,
    \label{eq:spherical-coords-general-lorentz}
\end{equation}
for some function $f(\varphi)$ determined by the warp factor $A(r)$. 
Because the causal structure does not depend on this unknown conformal factor, we again find the return locus to be 
\begin{equation}
    \rho = e^{i 2 \varphi_b}\;.
    \label{eq:return-radial-coord-general}
\end{equation}
where $\varphi_b = \int_{-\infty}^{r_b} e^{-A(r)} dr$. 
We note, in particular, that the causal structure of the bulk and of the return locus to the boundary is independent of the Euclidean distance to the brane. 
In contrast to this work, the Euclidean distance is what appears in holographic calculation of boundary entropy in 2D CFTs, for example, and many calculations of entanglement entropy. 

\paragraph{General geodesics}\label{sec:rgeneral-geodesics}
While calculating null geodesics which are not at a fixed position on the $\mathbb{S}^{d-2}$ would be slightly more challenging, there is no need to go to the trouble. 
The BCFT two-point function is a function only of a single cross-ratio, up to a conformally-covariant pre-factor. 
Thus, having mapped part of the null cone to the locus $\rho = e^{i 2 \varphi_b}$, we can conclude that null geodesics with non-zero momentum on the sphere must also return at another point on the sphere with the same cross-ratio.
Or, in other words, we can map any two unit vectors on the AdS$_d$ slices into each other by a conformal transformation and so all of the null rays are equivalent. 

\section{Looking for a Bulk Brane}\label{sec:bulk-brane}

In Section \ref{sec:simple-model} we showed how the bulk geometry of a simple ETW brane model is encoded in the spectrum and BOE coefficients of the dual BCFT. 
And in Section \ref{sec:lorentz} we showed that the bulk causal structure also predicts new Lorentzian singularities in the BCFT from null rays that reflect off the bulk ETW brane. 
We now put these two sides of the coin together and explain how one entails the other.\footnote{Or heads from one to the other. Pun obviously intended.} 

The boundary conformal block, written in terms of our radial cross-ratio \eqref{eq:rho-bblock}, has a simple large dimension limit
\begin{align}
    \lim_{\hat\Delta \rightarrow \infty} g_{\widehat{\Delta}}^{b}(\rho) 
    & = (4 \rho)^{\widehat{\Delta}} 
    \frac{1}{(1- \rho^2)^{(d-1)/2}}
    \; . \label{eq:rho-bblock-large-dim}
\end{align}
We consider this large-dimension limit at the return time of the bulk null cone, \eqref{eq:return-radial-coord}, to see that
\begin{align}
    \lim_{\hat\Delta \rightarrow \infty} g_{\widehat{\Delta}}^{b}(e^{i 2 \varphi_b}) 
    & = 4^{\widehat{\Delta}} e^{i 2 \widehat{\Delta} \varphi_b} 
    \frac{1}{(1- e^{i 4 \varphi_b})^{(d-1)/2}}
    \; . \label{eq:rho-bblock-large-dim-phase}
\end{align}
When we plug in the asymptotic spacing of boundary operator dimensions in our simple model, \eqref{eq:Delta_n-angular}, 
\begin{equation}
\widehat{\Delta} = \widehat{\Delta}_0 + n \frac{\pi}{\varphi_b}    
\end{equation}
we see that the block takes the form
\begin{equation}
     \lim_{n \rightarrow \infty} g_{\widehat{\Delta}_n}^{b}(\rho) 
    =  e^{i 2 \widehat{\Delta}_0 \varphi_b}  4^{\widehat{\Delta}_n}
    \frac{1}{(1- e^{i 4 \varphi_b})^{(d-1)/2}}\;.
    \, . \label{eq:aligned-blocks}
\end{equation}
Thus the spacing of the boundary operator dimensions has exactly cancelled the $n$-dependence of the phase precisely at the return time of bulk null cone. 
These conformal blocks will then all add coherently at this point so that the sum over conformal blocks takes the form
\begin{equation}
    e^{i 2 \widehat{\Delta}_0 \varphi_b}  4^{\widehat{\Delta}_0}
    \frac{1}{(1- e^{i 4 \varphi_b})^{(d-1)/2}} \sum_n 2^{2n\frac{\pi}{\varphi_b}}b_n
\end{equation}
Consequently, this sum can potentially diverge. 
To see that this is in fact the case, we plug in the large-$n$ BOE coefficients from \eqref{eq:boe-simple-model-large-n}. 
Dropping the prefactor, near the return time at $\rho = \exp\left[ i 2\varphi_b (1+\epsilon)\right]$ the sum over $n$ gives
\begin{equation}
\sum_n e^{2 \pi i n  \epsilon }n^{2\Delta-\frac{d+1}{2}}
\, .
\label{eq:divergent-sum}
\end{equation}
This is just a Fourier transform of the BOE coefficients. 
Doing the Fourier transform and extracting the singular contributions, we find Lorentzian singularities in the two-point function proportional to
\begin{equation}
    \mathcal{G}(\rho) \sim \frac{1}{(\rho_b - \rho)^{2\Delta -\frac{d-1}{2}}} \frac{1}{(-1 - \rho)^{\frac{d-1}{2}}}\, .
    \label{eq:toy-model-divergence}
\end{equation}

We conclude that the bulk causal structure has been mapped into a particular regular asymptotic spacing of the boundary operators that appear in the BOE. 

\paragraph{From bulk points to bulk branes }

Our story is a very close analogue, both in spirit and technically, to the story in \cite{Maldacena:2015iua}. 
There the authors explained how the causal structure of the dual AdS vacuum leads to new singularities in CFT four-point functions. 
These result from local interactions that happen at a point in the AdS bulk geometry. 
The bulk point isn't expected to be a true singularity of the four-point function---these are believed to occur only where predicted by Landau diagrams in the boundary theory. 
Rather it is a resonance in the correlator that is smoothed out at the scale of the cut-off where bulk locality breaks down. 

Similarly, we don't expect to find true new singularities in the BCFT two-point function. 
On the bulk side, we don't expect the brane to be exactly local. 
It will have some intrinsic width at which it will smear out bulk signals that reflect off the brane. 
On the boundary side, we only expect singularities where allowed by BCFT Landau diagrams. 
Thus, above some cutoff scale $\widehat\Delta_{\mathrm{gap}}$ that determines the width of the brane we expect the careful alignment of boundary operator dimensions to break down. 
Above this dimension, operators contribute with incoherent phases, truncating the divergent sum in \eqref{eq:divergent-sum}

\subsection{No bulk branes (at least generically)} \label{sec:no-bulk-branes}

We argued that we don't expect the bulk brane singularity to be a true singularity of the BCFT.
Nevertheless, the validity of our semiclassical description, a bulk geometry terminated by an ETW brane, over a large range of scales requires the careful alignment of boundary operator dimensions up to some large $\widehat\Delta_{\mathrm{gap}}$. 

We conjecture that this careful alignment is not a generic feature of BCFTs, even when the underlying CFT has a good gravitational description. Thus, an operator spectrum and BOE coefficients consistent with a bulk ETW brane geometry must be another input or assumption about the particular boundary condition of the CFT, much in the way we have to assume features of the spectrum of a large $c$ CFT such that it has a good semiclassical gravitational description. 

We do know that the correlation functions of a BCFT become those of the underlying CFT when all insertions are far from the boundary. 
Thus, we do not claim that the geometry will break down everywhere in the bulk. 
Rather, our claim is that generically there cannot be the type of simple causal structure consistent with an ETW brane geometry. 
The lack of fine-tuned dimensions prohibits null-rays from leaving the boundary and returning in reasonably-short times.\footnote{Of course, if we are willing to wait sufficiently long times, we can produce a resonance in an arbitrary theory by waiting for the phases of any finite number of blocks to align in the future. 
It's not clear that these types of resonances should have a \emph{simple} gravitational interpretation.}

In the spirit of \cite{Heemskerk:2009pn}, we can formalize our conjecture as the following:
\begin{conj}
A holographic CFT with boundary condition B will have a good bulk dual provided
\begin{enumerate}[(i)]
    \item correlation functions factorize about a (non-universal) free bulk solution; and
    \item the boundary operator spectrum is gapped such that below some large $\widehat\Delta_{\mathrm{gap}}$ the only operators are simple boundary operators and their multi-trace composites. 
\end{enumerate}
\end{conj}
It is the first of these conditions---the existence of a consistent leading-order free bulk two-point function---that we are concerned with in this paper and that we have argued shouldn't generically look like an ETW brane. 
Whether or not we might still expect a good bulk geometry, but without an ETW brane, we will discuss in section \ref{sec:beyond-etw-branes}. 
We leave the examination of the second of these conditions to future work, but we note that aspects of this are challenging without a characterization of the space of solutions to (i). 

Note that the causal structure of the four-point function in a holographic CFT also requires a similar alignment of operator dimensions. 
Specifically, in a holographic CFT one obtains ``double-twist operators'' due to the crossing equations and the presence of the identity operator; the stress tensor then fixes their anomalous dimensions, which asymptotically go to zero at both large spin and large central charge. 
These two facts explain the emergence of the ``bulk point'' \cite{Maldacena:2015iua}, where scattering between CFT operators occurs in the bulk but not the boundary.

In a defect CFT, a similar story generically emerges \cite{Lemos:2017vnx}: for a defect of codimension $q$, there are boundary operators associated to derivatives of bulk primaries in the $q$ directions transverse to the defect. Their anomalous dimension goes to zero at large ``transverse spin'' (i.e. the charge of the residual SO$(q-1,1)$ symmetry). 
This control of the anomalous dimensions clearly vanishes in an interface or boundary CFT. We no longer have any transverse spin to work with when $q=1$, even though we still have operators given by derivatives of bulk primaries in the remaining transverse direction. To have a good bulk dual, these operators must possess non-trivial anomalous dimensions that aren't fixed by symmetry and universal properties alone. A BCFT is then a simple setting where there isn't quite enough symmetry to fix the form of the vacuum two-point function and it must be an input. 


A useful analogy in holographic CFTs for when the free correlators are not fixed by symmetry is an excited state. 
Excited states in a holographic CFT will not generically have a good bulk geometry and hence will not have a good causal structure. 
Thus, we do not expect to see the approximate singularities of a local bulk geometry except in carefully chosen states. We suggest good ETW brane geometries are far from generic in the space of BCFTs in much the same way good bulk geometries are far from generic in the Hilbert space.

\section{Beyond ETW branes}\label{sec:beyond-etw-branes}

Our discussion so far has focused only on the bottom-up ETW brane proposal of \cite{Takayanagi2011}. 
However, we know this proposal is insufficient to fully describe various top-down models where the bulk geometry is more complicated. 
Moreover, while our evidence suggested that an ETW brane required finely-tuned boundary operator dimensions, we would like to be able to make a much more general conjecture: 
\emph{without special finely-tuned dimensions the BCFT bulk will not be everywhere geometric.} 
To examine this stronger statement, we will elaborate below on constructions that go beyond the ETW brane proposal. 

\subsection{General Bulk Geometries}\label{sec:sing-bulk-gen}

We explore first two simple toy models to build intuition for what happens when a BCFT is not terminated by an ETW brane, but terminates instead when extra dimensions pinch off. 

First, let us consider a metric on $HS^{2} \times \mathbb{R}^{0, 1}$, with the form
\begin{equation}
    ds^{2} = - dt^{2} + d \theta^{2} + \sin^{2} \theta d\varphi^{2} \: , \qquad t \in \mathbb{R} \: , \: \theta \in \left( 0, {\pi}/{2} \right) \: , \: \varphi \in (0, 2 \pi) \: .
    \label{eq:model-finite-causal-depth}
\end{equation}
We take this to be a model of an extra dimension pinching off in the IR bulk geometry where we view the angular direction $\varphi$ as an extra compact direction and $\theta$ as a coordinate on the base, like the bulk radial coordinate.  
We are concentrating on what happens in the region where the extra dimension pinches off and we ignore the rest of the geometry for now. 
Nevertheless, one could imagine smoothly joining the hemisphere to close off a circle of constant radius fibered over AdS. 

Suppose we have a massive scalar field $\phi$, with mass $M$, on this geometry; the equation of motion is
\begin{equation}
    M^{2} \phi = \Box \phi = - \partial_{t}^{2} \phi + \frac{1}{\sin \theta} \partial_{\theta} \left( \sin \theta \partial_{\theta} \right) \phi + \frac{1}{\sin^{2} \theta} \partial_{\varphi}^{2} \phi \: .
    \label{eq:model-finite-causal-depth-eom}
\end{equation}
We assume a simple boundary operator will correspond to some fixed momentum on the circle $\psi_m(\phi) = e^{i m \varphi}$. 
One might worry that the last term in \eqref{eq:model-finite-causal-depth-eom} will generate a potential for modes with non-zero angular momentum, leading to a different effective causal structure in the base space for different modes (ie. different boundary operators). More importantly, one might worry the potential will smear out the singularity of the returning signal that reflects off the end of the extra dimension. 
To see that this won't be the case, it's easiest to note that the causal structure corresponds to a high-energy limit where the potential becomes irrelevant---the pinching extra dimensions appear just like a hard wall when approached along the lightcone. 

We can also make the above intuition more explicit. 
For some fixed $m$, the solutions of the equation of motion that are smooth as the sphere caps off all take the form 
\begin{equation}
    \psi_{\omega,m}(t,\theta,\varphi) = P^{m}_{l(\omega)}(\cos\theta) e^{i\omega t} e^{i m \varphi} \, ,
\end{equation}
where $l(\omega) = \frac{1}{2}\left( \sqrt{4\omega^2-4M^2 +1}-1\right)$.  
Near the equator of the hemisphere, at large $\omega$, the solutions behave as
\begin{equation}
    \psi_{\omega,m}(t,\theta,\varphi) \approx \cos(\omega \theta+ \pi) e^{i\omega t} e^{i m \varphi}
\end{equation}
and so this behaves just like a wave reflecting off a hard wall at $\theta = 0$. 
The $\omega$ modes all add up coherently at $t = 2\theta$ to generate the reflected lightcone. 

We can further fix the allowed frequencies $\omega$ by imposing additional boundary conditions at the equator of the hemisphere (where we imagine joining onto the rest of the bulk solution). 
For example, we can set Neumann boundary conditions on the equator to find that the allowed asymptotic values of $\omega$ are precisely $\omega_n=2n\, , \; n\in \mathbb{N}$. 
It's then possible to see the lightcone that has reflected off of both boundaries and refocused at the original location: after a time inverse to the regular operator spacing the modes again add up coherently. 

Of course,  this isn't the only type of characteristic behaviour we can construct. 
We can also consider a toy model where the extra dimension caps off slowly:
\begin{equation}
    ds^{2} = - dt^{2} + dz^{2} + \frac{1}{z^2}d\varphi^{2} \: , \qquad t \in \mathbb{R} \: , \: z \in \mathbb{R}_{\geq z_0} \: , \: \varphi \in (0, 2 \pi) \: .
    \label{eq:model-infinite-causal-depth}
\end{equation}
Here the radius of the extra dimension shrinks slowly like $1/z$, again generating an effective potential for angular modes on the extra-dimension, but this time with no hard end. 
At the other end, we terminate the geometry at some $z_0$ (where again we could imagine joining onto another bulk solution). 
In contrast to the previous example, here we expect that as we increase the energy of modes they will penetrate deeper and deeper into the infrared geometry. 
Thus, there is no natural scale that sets the return time for lightrays sent in the radial direction and no expected reflected lightcone.

To see this explicitly, we can solve the wave equation for a free, massive scalar field in this background. 
The solutions that don't diverge in the infrared take the form
\begin{equation}
    \psi_{\omega,m}(t,z,\varphi) = U(\tfrac{M^2-\omega^2}{4 m},0;m z^2) e^{i\omega t} e^{i m \varphi} \, ,
\end{equation}
where $U(a,b,x)$ is the Tricomi (confluent hypergeometric) function. 
For large frequency (and sufficiently small $z$), we can rewrite these solutions as 
\begin{equation}
    \psi_{\omega,m}(t,z,\varphi) \approx \sqrt{\frac{2mz}{\pi \omega}}\cos(\omega z +\tfrac{\pi}{4}(1-\omega^2/m)) e^{i\omega t} e^{i m \varphi} \, .
\end{equation}
In contrast to the previous example, we see here that there is a phase shift that depends on $\omega$ and so we will not have all of the reflected modes add coherently at the same time/position. 

We can again further fix the allowed frequencies $\omega$ by imposing additional boundary conditions at $z_0$. 
For example, we can set Neumann boundary conditions at $z=1$ to find that the asymptotic spacing of eigenvalues $\omega_n$ scales like $\omega_n - \omega_{n-1} \sim 1/\sqrt{n} $. 
This spacing is not regular, and thus we will not have a finite-time recurrence where a reflected lightcone could return. 

\paragraph{Causal Depth}

The primary distinguishing feature of the above two examples is the difference in their \emph{causal depth}.
As in the ETW brane examples, when light rays sent into the geometry can return in finite time (a finite causal depth), we see a reflected lightcone, a corresponding divergence in the two-point function, and the careful alignment of asymptotic eigenvalues. 
On the other hand, when light rays sent into the geometry don't return in finite time, we no longer expect a reflected lightcone, nor a corresponding divergence in the two-point function or alignment of operator dimensions. 
The common intuition is that BCFTs will have a dual bulk geometry with a finite causal depth, whether they are terminated by an ETW brane or the closing off of extra compact dimensions, and should look more like the first of these very simple toy models.

\subsubsection{Asymptotically AdS Toy Model} \label{sec:AdStoymodel}

The comments above contain the essential features and physical intuition for the argument that individual Kaluza-Klein modes exhibit the same reflected singularity (in the case of finite causal depth), and consequently, that different CFT operators must exhibit the same resultant Lorentzian singularity.\footnote{Although this argument is clearest when the form of the warping is relatively simple, as in the examples we have described, we might hope that something analogous should hold in any case where there are internal directions that degenerate. }
To give additional support for this claim, we can explicitly analyze another toy model involving a free scalar field propagating on a background which is a static, asymptotically locally AdS geometry whose spatial slices are cigar-like, as described by the metric
    \begin{equation}
    ds^{2} = \frac{L^{2}}{z^{2}} \left( dz^{2} - dt^{2} + z^{2} f(z)^{2} d \theta^{2} \right) \: , \qquad \theta \in (0, 2 \pi) \: ,
    \end{equation}
    where for convenience we choose
    \begin{equation}
        f(z) = \frac{1 - z^{2}}{2} \: .
    \end{equation}
    One can verify that this choice makes the geometry non-singular at $z = 1$. 

\begin{figure}
    \centering
    \includegraphics[height=7cm]{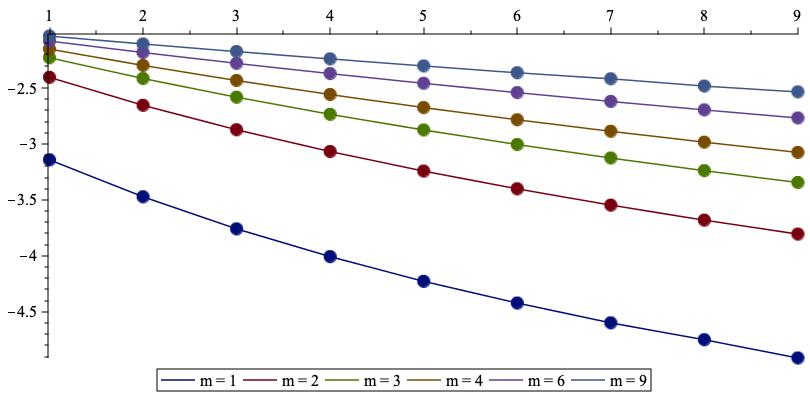}
    \caption{Plots of $\ln \big| \omega_{n+1} - \omega_{n} - \pi \big|$ versus $n$ for various values of $m$ in asymptotically AdS toy model with a free scalar field. We have chosen $ML = 1$ for concreteness.}
    \label{fig:AdSspacingexample}
\end{figure}

The explicit analysis of this problem is left to Appendix \ref{app:AdStoymodel}, but we summarize the important features here.
The eigenfunctions of the Klein-Gordon operator are of the variable-separated form $e^{i m \theta} e^{i \omega t} \phi_{m, \omega}(z)$, where we interpret different KK modes of the internal $S^{1}$ (labelled by integer $m$) as corresponding to different operators in the dual quantum theory. The functions $\phi_{m, \omega}(z)$ are known (they involve linear combinations of \textit{confluent Heun functions}, with linear dependence fixed by normalizability at $z=0$), and the values of $\omega$ are quantized by the requirement of regularity at $z=1$. 
To demonstrate that 
operators of differing $m$ exhibit the same reflected singularity, which we expect to occur when the insertions are separated by $\Delta t  =2$ (the time for an ingoing null ray to travel from $z=0$ to the IR wall at $z=1$ and bounce back), it suffices to show that, for any fixed $m$, the quantized frequencies exhibit asymptotic spacing
\begin{equation} \label{eq:asymptspacing_AdStoymodel}
    \Delta \omega \equiv \lim_{n \rightarrow \infty} \omega_{n+1} - \omega_{n} = \frac{2 \pi}{\Delta t} = \pi \: .
\end{equation}
The form of the Green function as a sum over modes then fixes the desired singularity.
Computing the spectrum numerically in various cases, we indeed appear to find that (\ref{eq:asymptspacing_AdStoymodel}) holds across values of $m$; an example of this calculation can be seen in Figure \ref{fig:AdSspacingexample}.
We interpret this as additional concrete evidence for our claim that CFT operators corresponding to KK modes of some bulk field should quite generally exhibit a  common Lorentzian singularity associated with bulk causality.

\subsection{Top Down Holographic BCFT}\label{sec:top-down}

In the previous subsection, we have discussed how some of our conclusions may be generalized to plausible smooth bulk geometries involving an internal space which degenerates. We would now like to understand how these considerations might apply to fully-fledged top-down holographic constructions of AdS/BCFT. As mentioned in Section \ref{sec:rev-holo-top}, a number of such constructions are known 
\cite{Chiodaroli:2011fn, Chiodaroli:2012vc, Berdichevsky:2013ija, Bachas:2013vza, DHoker:2007zhm, DHoker:2007hhe, Aharony:2011yc}, and it appears necessary to explain whether or not the existence of a smooth bulk in these cases entails the spectral fine-tuning we have argued for previously, and if not, how their causal structure is consistent with this conclusion. Since the spectrum of boundary operators with protected conformal dimensions is not known in these theories, and extracting this spectrum is not expected to be straightforward in general \cite{Bachas:2011xa}, our investigation will be on the side of the bulk causal structure. 

Certainly, it is a possibility that the heuristic arguments of the previous subsection may not capture important structural aspects of the warped supergravity solutions describing the microscopic theories of interest. But in this section, we will attempt to address a different possibility, namely that the intersection of the bulk light cone of a boundary point with the asymptotic boundary may generically be small or empty in these theories. In this case, bulk causality would not directly introduce a spectral constraint of the type we have been concerned with, or would do so only for a very select subset of operators. In the language of the previous subsection, we might say that such solutions exhibit infinite causal depth. Unlike the simple toy models that we have considered, this is no longer necessarily a manifestation of a slowly shrinking internal space, but rather may reflect that a generic ingoing geodesic will follow a trajectory which may involve very complicated behaviour in the internal space. 

We can illustrate this possibility by considering a concrete example,
namely the ``AdS$_{2}$-Cap" solution of six-dimensional $(0, 4)$ (``type 4b") supergravity identified by Chiodaroli, D'Hoker, and Gutperle in \cite{Chiodaroli:2011fn}; this is expected to provide a fully back-reacted holographic description of a D1-D5 junction in 10 dimensions, where a D1-brane and a D5-brane wrapping a suitable four-manifold join to form a D1/D5 bound state. This solution has a single AdS$_{3} \times S^{3}$ asymptotic region, describing the near-horizon geometry for the D1/D5 bound state, while the D1-brane and the D5-brane each contribute a curvature singularity at fixed locations in the internal space, referred to as ``caps". The resultant 6-dimensional metric is of the form
\begin{equation}
    ds_{6}^{2} = f_{1}^{2} ds_{\textnormal{AdS}_{2}}^{2} + f_{2}^{2} ds_{S^{2}}^{2} + \rho^{2} dw d\bar{w}  \: ,
\end{equation}
where $(w, \bar{w})$ are complex coordinates on a Riemann surface $\Sigma$, which we may take to be the upper half plane. One may use $SL(2, \mathbb{R})$ symmetry to fix the location of the AdS$_{3} \times S^{3}$ asymptotic region at $w = 0$ and the AdS$_{2}$-caps at $w = \pm 1$.

\begin{figure}
\centering
\begin{subfigure}{.45\textwidth}
  \centering
  \includegraphics[height=6.5cm]{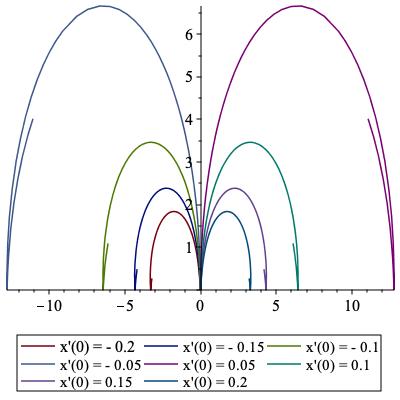}
  \caption{Initial behaviour of ingoing null geodesics on internal submanifold $\Sigma$ (with initial data in the $x$-direciton provided).}
  \label{fig:m2s}
\end{subfigure} 
\quad
\begin{subfigure}{.45\textwidth}
  \centering
  \includegraphics[height=6.5cm]{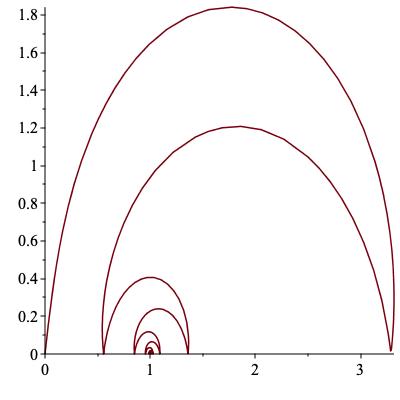}
  \caption{Generic behaviour of ingoing null geodesic (in this case $x'(0)=0.2$), which falls into and orbits the AdS-cap. }
  \label{fig:m1s}
\end{subfigure} 
\caption{The AdS-cap curvature singularities at $w = \pm 1$ in the upper half-plane $\Sigma$ appear to be attractor-like for ingoing null geodesics.}
\label{fig:AdSCapGeodesics}
\end{figure}

Equipped with this solution, one can proceed to study numerically the behaviour of ingoing null geodesics; 
we leave a more detailed discussion to Appendix \ref{app:topdown}. In Figure \ref{fig:AdSCapGeodesics} we display the behaviour of such null geodesics on the internal submanifold $\Sigma$, restricting to geodesics with no initial momentum in the $S^{2}$ direction for clarity; the geodesics are seen to initially orbit the caps, which appear to be attractor-like.

To attempt to confirm that the AdS-caps do indeed attract all ingoing null geodesics in the class we are considering, we can examine the geodesic equation in the vicinity of these caps, again restricting to the case with no momentum in the $S^{2}$ direction. (We choose the one at $w=1$ for convenience). Denoting
\begin{equation}
    x-1 \equiv \epsilon_{x} \equiv \epsilon \cos \Theta \: , \qquad y \equiv \epsilon_{y} \equiv \epsilon \sin \Theta \: ,
\end{equation}
we obtain at leading order in small $\epsilon$
\begin{equation}
        \frac{\ddot{\epsilon}}{\epsilon} + \frac{\dot{\epsilon}^{2}}{\epsilon^{2}} 
        = 2 \frac{\dot{\epsilon}^{2}}{\epsilon^{2}} +  \left( O(\epsilon)  \frac{\dot{\epsilon}_{x}^{2}}{\epsilon^{2}} +   O(\epsilon) \frac{\dot{\epsilon}_{y}^{2}}{\epsilon^{2}} + O(\epsilon) \frac{\dot{\epsilon}_{x}}{\epsilon} \frac{\dot{\epsilon}_{y}}{\epsilon} \right) \: ,
\end{equation}
from which it should follow that
\begin{equation}
    \epsilon(s) \approx \epsilon_{0} e^{c s} \: ,
\end{equation}
provided that $\epsilon$ remains sufficiently small 
that perturbation theory remains valid. That is, it appears that the geodesics will approach the AdS-cap exponentially, and in particular would not reach the precise location of the cap in finite affine parameter in the leading order calculation. We can confirm numerically that, at least for sufficiently small affine parameter, this is a good approximation (see Appendix \ref{app:topdown}).

Based on this cursory analysis, it appears plausible that a significant portion of the bulk light cone has no intersection with the asymptotic boundary of the spacetime, instead getting trapped near the AdS-cap curvature singularities. It is in this sense that this example may be exhibiting what we have previously referred to as an infinite causal depth. Note that, in this case, this is not due to a slowly shrinking internal space; in fact, the AdS-caps are at finite proper distance from any point in the interior of the geometry, and the geometry in the vicinity of the AdS-cap is of the form
\begin{equation}
\begin{split}
    ds^{2} & \approx \frac{\sqrt{2} \mu_{0}}{\kappa} \sqrt{ \kappa^{2} + \mu_{0}^{2} \sin^{2} \Theta} \Bigg( d \epsilon^{2} +  \epsilon^{2} \left( d \Theta^{2} + \frac{\kappa^{2}}{\mu_{0}^{2}} ds_{\textnormal{AdS}_{2}}^{2} + \frac{\kappa^{2} \sin^{2} \Theta}{\kappa^{2} + \mu_{0}^{2} \sin^{2} \Theta} ds_{S^{2}}^{2} \right) \Bigg) \: , 
\end{split}
\end{equation}
where $\kappa > 0$ and $\mu_{0}$ are the real parameters for our class of AdS$_{2}$-cap solutions. 
Rather, this is simply a seemingly generic property of null geodesics sent inward from the asymptotic boundary.

In general, we expect that the supergravity approximation should break down in the vicinity of these singularities, with string loop and $\alpha'$-corrections becoming important. This may also be a generic feature of top-down constructions of holographic BCFT: generic causal probes will enter regions of the bulk where the effective description breaks down. 

It is worth noting that, even in this example, there exist geodesics which return to the asymptotic boundary in short affine parameter as a result of finely-tuned initial conditions. For example, null geodesics sent radially inward from the equator of the asymptotic $S^{3}$ 
will avoid falling into either AdS-cap. In the BCFT, the radial coordinate cross-ratio describing the separation between endpoints of these geodesics is precisely
\begin{equation}
    \rho = e^{\pm \frac{i \pi \mu_{0}}{2 \kappa}} ;.
\end{equation}
(See Appendix \ref{app:topdown} for further details.) It is possible that this result is indicating the existence of more complicated operators localized on the internal dimensions which do exhibit a singularity, and that this again requires the existence of particular families of boundary operators (whose dimensions in this case have asymptotic spacing $|\Delta_{n+1} - \Delta_{n}| \sim \frac{4 \kappa}{mu_{0}} n$). 

\subsubsection{Top-down models with finite causal depth}\label{sec:top-down-finite}

The above top-down BCFT example shows how the holographic dual can seem to develop an infinite causal depth and avoid the simplest causal constraints on the BOE spectrum. 
Here we show that this isn't the case for all known top-down constructions: 
we give an explicit example of a top-down BCFT with finite causal depth. 

Our example will be an interface conformal field theory (ICFT) with a codimension one defect separating two different CFTs on either side of the defect. 
We can view the ICFT as a BCFT by folding the two sides on top of each other. 
Then the BCFT will look like a product of two non-interacting theories that are coupled only via the boundary condition. 

The precise example we will consider here is the supersymmetric Janus solution found in \cite{Chiodaroli:2010ur}. 
In this Janus solution, the bulk geometry smoothly interpolates between two different asymptotic AdS$_3\times S^2 \times T^4$ regions. 
The dual ICFT is a marginal deformation of the two-dimensional D1/D5 $\mathcal{N} = (4,4)$ SCFT. 
The details of the geometry will not be important to our analysis and can be found in \cite{Chiodaroli:2010ur}. 
For our purposes, we note that the metric takes a form similar to the previously discussed top-down constructions:
\begin{equation}
d s_{10}^{2}=f_{1}^{2} d s_{A d S_{2}}^{2}+f_{2}^{2} d s_{S^{2}}^{2}+\rho_{}^{2} d w d \bar{w}+f_{3}^{2} d s_{T^{4}}^{2}\;,
\end{equation}
where $w,\bar w$ are coordinates on a Riemann surface $\Sigma$ with boundary. 
The various metric prefactors are functions on the Riemann surface. 

Let us note that $\kappa$ is bounded, $\kappa>1$, and controls the warping of the AdS$_2$ slices of the bulk. 
In \cite{Chiodaroli:2010ur}, it was shown that the boundary entropy is determined in terms of $\kappa$ and takes the form
\begin{equation}
    S_{bdy} = \frac{c}{3}\log \kappa \, ,
\end{equation}
where $c$ is the central charge of the 2D CFT. 
With the bound on $\kappa$ the boundary entropy is always positive.

In \cite{Chiodaroli:2016jod}, the Euclidean two-point function was computed holographically and takes the form
\begin{equation}
\begin{aligned}
\left\langle\mathcal{O}\left(x_{\perp}, t\right) \mathcal{O}\left(x_{\perp}^{\prime}, t^{\prime}\right)\right\rangle=& \frac{2^{3-2 \Delta}}{\pi \kappa^{2 \Delta}[\Gamma(\Delta-1)]^{2}} \frac{1}{\left|x_{\perp} x_{\perp}^{\prime}\right|^{\Delta}} \times \\
& \times \sum_{n=0}^{\infty} \operatorname{sign}(\xi)^{n}(n+\Delta-1 / 2) \frac{\Gamma(n+2 \Delta-1)}{\Gamma(n+1)} Q_{\widehat \Delta_n -1}(|2 \xi+1|) \, .
\end{aligned}
\label{eq:interface-2pt}
\end{equation}
The boundary operator dimensions $\widehat \Delta_n$ appearing in the calculation were found exactly in \cite{Chiodaroli:2016jod} and their asymptotic spacing is given by
\begin{equation}
    \lim_{n\rightarrow\infty} \widehat \Delta_n - \widehat \Delta_{n-1} = \frac{1}{\kappa} \, .
\end{equation}
From this careful alignment of boundary operator dimensions we can already conclude that these solutions have a finite causal depth. 

To see the finite causal depth explicitly, we can go ahead and find the bulk singularity by looking at the asymptotics of the analytically continued two-point function. Again using our radial coordinates, $\rho$, we expand \ref{eq:interface-2pt} when both points are spacelike separated on the same side of the defect. 
As already noted in \cite{Chiodaroli:2016jod}, this gives an expansion in terms of $\rho^{n/\kappa}$.\footnote{In the original author's notation, our radial coordinates is precisely their $\zeta$.} Thus, when we analytically continue to the timelike region on the RHS where $\rho = e^{i \theta_R}$, we expect a singularity at a time
\begin{equation}
    \theta_R = 2\pi \kappa \, .
\end{equation}
This can be explicitly checked from the summation of the asymptotic expansion in \cite{Chiodaroli:2016jod}, which gives a singularity of the form
\begin{equation}
    \frac{1}{(1-\rho^{1/\kappa})^{2\Delta-1/2}} \, ,
\end{equation}
with the appropriate divergence at $\rho = e^{i 2\pi \kappa}$. 

We can also calculate the expansion when the two operators are timelike separated on opposite sides of the defect. 
We first insert the second operator in the spacelike region across the defect where $0<-\rho<1$. 
Here the expansion takes the schematic form
\begin{equation}
    \sum_n  c_n (-1)^n (-\rho)^{n/\kappa} \, .
\end{equation}
We can then analytically continue $-\rho$ to live on the unit circle and find that the phases align when
\begin{equation}
    \theta_L = \pi \kappa \, .
\end{equation}
As expected, we find the singularity on the LHS of the defect at half the time it takes for it to return to the RHS.

\section{Discussion}\label{sec:discussion}

We have argued that a powerful probe of the putative bulk geometry of a BCFT is the Lorentzian two-point function. 
The two-point function is sensitive to the (approximate) causal structure of the bulk and is a probe of how null rays can reflect off the IR geometry and return to the boundary.

In the case that the bulk geometry is terminated by an ETW brane, we argued in Section \ref{sec:simple-model} that this is indicated in the two-point function of simple CFT operators by a fixed, careful alignment of the boundary operator dimensions appearing in the BOE. 
We suggest that there is no reason to expect such spacing generically in the possible boundary conditions for a given holographic CFT. 
Thus, we have argued that an ETW brane is not generically the correct bulk description of a boundary condition for a holographic CFT. 

In certain cases, we also have a top-down construction of the dual geometry for a boundary condition of a holographic CFT. 
These geometries are not described by ETW brane geometries, but rather have extra compact dimensions that pinch off in the IR. 
We argued that simple geometries that cap off at a finite causal depth will also have a reflected lightcone that relies on a seemingly fragile alignment of boundary operator dimensions. 
However, in some of the more complicated geometries actually found in top-down constructions, it is more subtle. 
There it seems that many null geodesics may become trapped in the IR geometry and do not return at short times. 
These top-down constructions then might be more similar to our toy models without a finite causal depth. 

The existence of top-down constructions with both finite and infinite causal depth make it less clear just how atypical the existence of a good bulk dual is for a BCFT. 
Because some examples seem to not have a finite causal depth when probed by simple CFT operators, they do not require the same careful alignment of boundary operator dimensions. 
Nevertheless, it would be a much stronger claim to then assume that any set of boundary dimensions could always be explained by a sufficiently-complex local bulk geometry. 
It seems more likely that these complicated geometries would be a small subset of even more complicated, non-local solutions. 
Moreover, by probing a geometry with more complicated operators that are localized in the extra dimensions, we might expect to be able to send light-like signals into the bulk that return to the boundary. 
These would then require constraints on the boundary spectrum of these more complicated operators, which also need not be generic in the space of boundary conditions. 

\paragraph{Complex geometries and chaotic boundary spectra}
The trapping of light-like geodesics in the IR geometry of some top-down models bears a  resemblance to the dynamics of chaotic systems.
Insofar as the null geodesics explore the IR geometry ergodically and fail to return to the boundary, this picture is reminiscent of the chaotic motion of billiards on a Reimannian manifold. 
One possible correspondence is that there is a mapping between the irregular spectrum of light operators in the BCFT and the chaotic light-like trajectories in the bulk geometry. 
This seems to be a different manifestation of chaotic dynamics than usually discussed in holography and it would be interesting to understand this connection better\footnote{The possibility that the ``hard chaos" of such causal probes in the near-horizon region of a black hole could be a manifestation of scrambling appeared in \cite{Barbon:2011pn}. 
}. 
Nevertheless, as stated above, we remain hesitant to suggest that any complicated spectrum could be mapped to a sufficiently complicated \emph{local} bulk geometry. 

\paragraph{Boundary vs. bulk causality}

In the ETW brane scenario, when the brane is close to the boundary, bulk null rays can reflect off the bulk brane and return to their point of origin more quickly than a null ray confined to the boundary. 
This is the region (in Figure \ref{fig:lorentzian-causal-structure}) where $-1< \xi < 0$. 
In a 2D CFT with the simplest AdS+ETW brane bulk, for example, this happens when the boundary entropy is negative. 

There is some apparent (if perhaps naive) tension here with causality: a bulk observer can learn information about the boundary condition more quickly than they can causally probe the boundary of the CFT itself. 
On the other hand, these signals return in the causal future of the boundary point, so there is no sharp conflict with boundary causality. 
Moreover, it's important to note that information about the boundary condition isn't localized at the boundary itself. 
As just one obvious example, information about the boundary condition is encoded in one-point functions measurable arbitrarily far from the boundary. 

There are other cases where a bulk singularity in the region $-1< \xi < 0$ would actually be in conflict with boundary causality. 
In an ICFT (folded to be seen as a BCFT) a bulk singularity in this region between a RHS and LHS operator would correspond to a signal travelling acausally across the defect to the other side. 
We note that in Section \ref{sec:top-down-finite}, our Janus solution did not have a singularity in this region precisely because $\kappa>1$ (that is the boundary entropy was greater than zero). 

It would be interesting to have top-down constructions where information about the boundary can be causally accessed more quickly via the bulk than via the CFT.

\paragraph{Top-down models and SUSY}
It is believed by some that supersymmetry is a necessary ingredient for the existence of holographic CFTs \cite{Ooguri:2016pdq}. 
In our top-down constructions, it would be interesting to know what role, if any, supersymmetry plays in fixing the boundary operator spectrum so that it is consistent with the bulk geometry. 
This would be particularly interesting to consider in the ICFT of Section \ref{sec:top-down-finite} where we do see the careful tuning of asymptotic boundary operator dimensions. 

\paragraph{Calculating BCFT two-point functions}
 As we discussed in Section \ref{sec:top-down}, we do have top-down constructions of holographic BCFTs with strong evidence for the existence of a good bulk geometry. 
 It would be useful to have BCFT calculations of two-point functions in these theories to confirm the correspondence with predictions from the bulk geometry.  

\paragraph{Bootstrap constraints}
We have argued that the constraints on the boundary spectrum necessary for agreement with a simple bulk geometry appear fragile and are not expected to be generic. 
Moreover, the existence of more complicated top-down constructions also seems to imply that a fixed regular spacing cannot be the only allowed possibility. 
Nevertheless, we have not ruled out the possibility that the alignment of boundary operator dimensions follows from some simpler assumptions, perhaps by using an appropriate bootstrap argument. 
It would be interesting to explore this further.

\paragraph{2D CFTs}

In \cite{Sully:2020pza}, it was shown that the entanglement entropy of an interval in a 2D BCFT is consistent with the AdS+ETW brane proposal, provided the assumption of vacuum block dominance in the BCFT. 
It is somewhat surprising that the bulk would reproduce the correct entanglement entropy, even if it fails to satisfy the constraints laid out in this paper. 
One possible resolution is that the entanglement entropy is a rather weak probe of the bulk geometry in this setting. 
When in this disconnected phase, the entropy depends only on the boundary entropy and measures only the integrated distance to the brane. 
On the other hand, it would be interesting if the assumption of vacuum block dominance also placed constraints in the Lorentzian bulk brane regime we considered here. 


\acknowledgments

We thank Alexandre Belin, Arjun Kar, Andreas Karch, Lampros Lamprou, Eric Perlmutter, Tadashi Takayanagi, and Mark Van Raamsdonk for useful discussions and/or comments on the draft. This work is supported by the Natural Sciences and Engineering Research Council of Canada (NSERC). DW is supported by an IDF scholarship from UBC. CW is supported by the NSERC PGS-D program. 



\appendix

\section{Bulk Canonical Commutators}\label{app:bulk-field-commutators}
In section \ref{sec:rev-holo-dict} we showed that a bulk scalar operator $\phi$ can be expressed as
\begin{equation}\label{eq:field_expansion}
    \phi(\vec y, u, r) = \sum_n c_n \bar\psi_n(r) \hat\phi_n(\vec y ,u) \,, 
\end{equation}
where the $\hat\phi_n(\vec y,u)$ are codimension-1 fields of asymptotic dimension \eqref{eq:Delta_n}, and $\bar\psi_n(r)$ are the normalizable mode functions (normalized with leading unit coefficient near the asymptotically-AdS boundary). 
That analysis was insufficient to determine the overall constants $c_n$. 

To determine the $c_n$ we require that the operators $\phi$ and $\hat\phi_n$ satisfy the equal time canonical commutation relations
\begin{align}
    \left[ \phi(t, \vec v_1, u_1, r_1) , \pi(t, \vec v_2, u_2, r_2) \right] &= 
    i \delta^{d+1}(\vec v_1 - \vec v_2, u_1-u_2, r_1 - r_2),\\
    \left[ \phi_n(t,\vec v_1,u_1), \pi_m(t,\vec v_2,u_2) \right] 
    &= 
    i \delta^{d}(\vec{v}_1-\vec{v}_2,u_1-u_2) \delta_{m,n},
\end{align}
where $\vec y = (t,\vec v)$ and where $\pi\equiv -\sqrt{-g}g^{tt}\partial_t \phi$ and $\pi_n\equiv -\sqrt{-\tilde g}\tilde g^{tt}\partial_t \phi_n$ are the canonically conjugate fields to $\phi$ and $\pi_n$, with $\tilde g$ the induced metric on the AdS${}_d$ slices of fixed $r$. This gives
\begin{equation}
    \int_{r_0}^\infty dr \cosh^{d-2}(r)\psi_n(r)\psi_m(r) = \frac{1}{c_n^2} \delta_{nm},
\end{equation}
which we can evaluate to obtain $c_n$. 

\section{Asymptotically AdS Toy Model and the Reflected Singularity} \label{app:AdStoymodel}

In this appendix, we will provide some details for the calculation involving a free scalar field propagating on an asymptotically locally AdS background that was mentioned in Section \ref{sec:AdStoymodel}. Recall that our goal is to provide evidence in a concrete example that different KK modes will give rise to the same reflected singularity when the bulk geometry has finite causal depth. As pointed out earlier, it is sufficient to demonstrate that the asymptotic frequency spacing $\Delta \omega$ for each mode in this model is equal to $\pi$; we approach this problem numerically. 

Recall that the background geometry is
\begin{equation}
    ds^{2} = \frac{L^{2}}{z^{2}} \left( dz^{2} - dt^{2} + z^{2} f(z)^{2} d \theta^{2} \right) \: , \qquad f(z) = \frac{1 - z^{2}}{2} \: , \qquad \theta \in (0, 2 \pi) \: .
\end{equation}
The massive scalar wave equation $\Box \phi = M^{2} \phi$ on this background yields
\begin{equation}
    \partial_{z}^{2} \phi - \frac{2z}{1 - z^{2}} \partial_{z} \phi = \left( \partial_{t}^{2} - \frac{4}{z^{2} (1 - z^{2})^{2}} \partial_{\theta}^{2} + \frac{M^{2} L^{2}}{z^{2}} \right) \phi \: .
\end{equation}

Separating variables by writing
\begin{equation}
    \phi(z, t, \theta) =  e^{i m \theta} e^{i \omega t} \phi_{m, \omega}(z) \: , 
\end{equation}
the $\phi_{m, \omega}(z)$ must satisfy
\begin{equation}
    \partial_{z}^{2} \phi - \frac{2z}{1 - z^{2}} \partial_{z} \phi = \left(- \omega^{2} + \frac{4 m^{2}}{z^{2} (1 - z^{2})^{2}} + \frac{M^{2} L^{2}}{z^{2}} \right) \phi  \: .
\end{equation}
This equation has general solution
\begin{equation}
    \begin{split}
        \phi(x) & = c_{1} (z^{2} - 1)^{m} x^{\frac{1}{2} + \frac{\sqrt{1 + 4 M^{2} L^{2} + 16 m^{2}}}{2}}   H\left( 0, \frac{\sqrt{1 + 4 M^{2} L^{2} + 16 m^{2}}}{2}, 2m, \frac{\omega^{2}}{4}, \frac{8 m^{2} - \omega^{2} + 1}{4}, z^{2} \right) \\
        & + c_{2} (z^{2} - 1)^{m} x^{\frac{1}{2} - \frac{\sqrt{1 + 4 M^{2} L^{2} + 16 m^{2}}}{2}}   H\left( 0, - \frac{\sqrt{1 + 4 M^{2} L^{2} + 16 m^{2}}}{2}, 2m, \frac{\omega^{2}}{4}, \frac{8 m^{2} - \omega^{2} + 1}{4}, z^{2} \right) \: ,
    \end{split}
\end{equation}
where $H(\cdot)$ is a confluent Heun function, and $c_{1}, c_{2}$ are undetermined constants. The Heun function is analytic at the point where its last argument vanishes (the asymptotic boundary $z = 0$), where it takes value $H(\ldots, 0) = 1$, and is singular for generic $m, \omega$ when its last argument is unity (the ``IR wall" $z = 1$). For $M^2 > 0$, we see that the first function is the normalizable solution at $z=0$; for negative $M$ above the Breitenlohner-Freedman bound
\begin{equation}
    - 1 \leq M^{2} L^{2} < 0 \: ,
\end{equation}
we also have an alternate quantization as usual. 

\begin{figure}
\centering
\includegraphics[height=7.5cm]{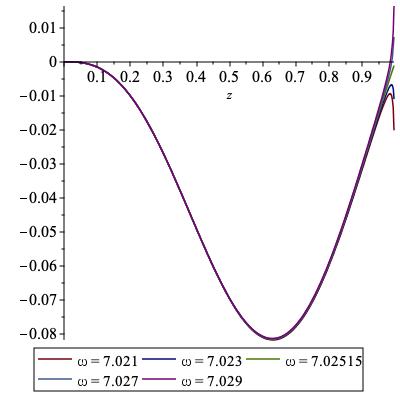} \quad
\includegraphics[height=7.5cm]{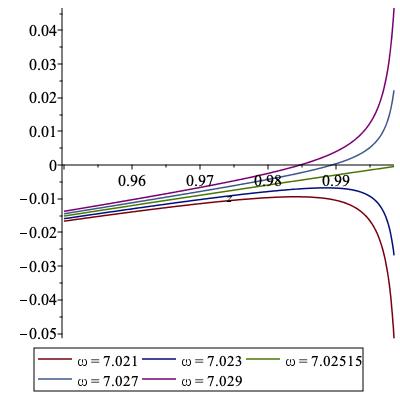}
\caption{Illustration of procedure for determining allowed frequencies $\omega$. We plot solutions $\phi_{m, \omega}(x)$ with $m=1$ consistent with normalizability at $z=0$, for various values of $\omega$. We have chosen $ML = 1$ for concreteness. The figure on the left shows the full range of $z$, while the figure on right shows a close-up in the vicinity of $z=1$. }
\label{fig:AdS_approach}
\end{figure}

With the general solution to the equation of motion in hand, we will proceed to fix various values of $m$, and numerically determine the values of $\omega$ consistent with regularity at $z=1$. The appropriate boundary condition to impose is
\begin{equation}
    \phi_{m, \omega}(z) \stackrel{z \rightarrow 1}{\sim} (1-z)^{m} \: ,
\end{equation}
since otherwise the $\theta$-dependence will generate a singularity at $z=1$. 
When computing the allowed frequencies numerically, we simply look for roots of $\phi_{m, \omega}(z=1)$ as a function of $\omega$. In particular, we do not manually impose the correct power law $(1-z)^{m}$ in the vicinity of $z=1$, but rather we merely require that $\phi_{m, \omega}(z=1)$ vanishes. For $m \neq 0$ and generic $\omega$, $\phi_{m, \omega}(z)$ actually has a singularity at $z=1$; the desired values of $\omega$ occur when $\phi_{m, \omega}$ transitions between becoming large and positive and becoming large and negative in the vicinity of this singularity. 
This procedure is illustrated in Figure \ref{fig:AdS_approach}.

Applying this procedure for various values of $m$, we indeed find that the spacing of allowed frequencies $\omega$ quickly converges to $\Delta \omega = \pi$, precisely as required to produce the singularity at $\Delta t = 2$; see Figure \ref{fig:AdSspacingexample} of Section \ref{sec:AdStoymodel}. We note that this convergence appears to become slower with increasing $m$, though there is no reason to expect that the convergence would break down at any finite $m$.

\section{BCFT Singularities in 2D}\label{app:2d-sing}


Consider insertions $z_i$ on the (Euclidean) upper-half plane, with distances $z_{ij}$ and $z_{i\bar{j}}$ defined as usual.
We will be interested in the correlator $\langle \calO(z_1)\calO(z_2)\rangle$ and its Lorentzian continuation.

\subsection{OPE expansion}

The OPE expansion of our BCFT two-point correlator is just a sum over holomorphic Virasoro conformal blocks.
In the bulk CFT channel $\xi \to 0$, we have
\begin{equation}
    \mathcal{F}(\eta) := \langle \calO(z_1)\calO(z_2)\rangle = \sum_h C_{\calO\calO h} A_h \mathcal{V}_h(1-\eta),
\end{equation}
where $C_{\calO\calO h}$ are bulk CFT OPE coefficients, $B_h$ is the one-point function associated with the primary $h$, and $\xi = (1-\eta)/\eta$.
In the boundary channel $\eta \to 0$ ($\xi \to \infty$), we have
\begin{equation}
    \mathcal{F}(\eta) = \sum_{\hat{h}} |B_{\calO \hat{h}}|^2 \mathcal{V}_{\hat{h}}(\eta),
\end{equation}
where $B_{\calO h}$ is a boundary OPE coefficient.
It is possible to expand Virasoro blocks as \cite{recurse}
\begin{align}
    \mathcal{V}_h(\eta) & = (16q)^{h-(c-1)/24}[\eta(1-\eta)]^{(c-1)/24-2h}\theta_3(q)^{(c-1)/2-16h}H(h, q),
    \label{eq:Vir-block}
\end{align}
where $H(h, q)$ is a power series in $q$ which can be determined recursively, $\theta_3$ is a Jacobi theta function, and $q$ is the elliptic nome defined by 
\begin{align}
    q & = e^{i\pi \tau(\eta)}, \quad
    \tau = i \frac{K(1-\eta)}{K(\eta)}, \quad K(\eta) = \frac{1}{2}\int_0^1 \frac{dt}{\sqrt{t(1-t)(1-\eta t)}}.
\end{align}
This can be inverted to give $\eta = [\theta_2(q)/\theta_3(q)]^4$.

\subsection{The pillow geometry}

\begin{figure}
    \centering
    \includegraphics[width=0.5\textwidth]{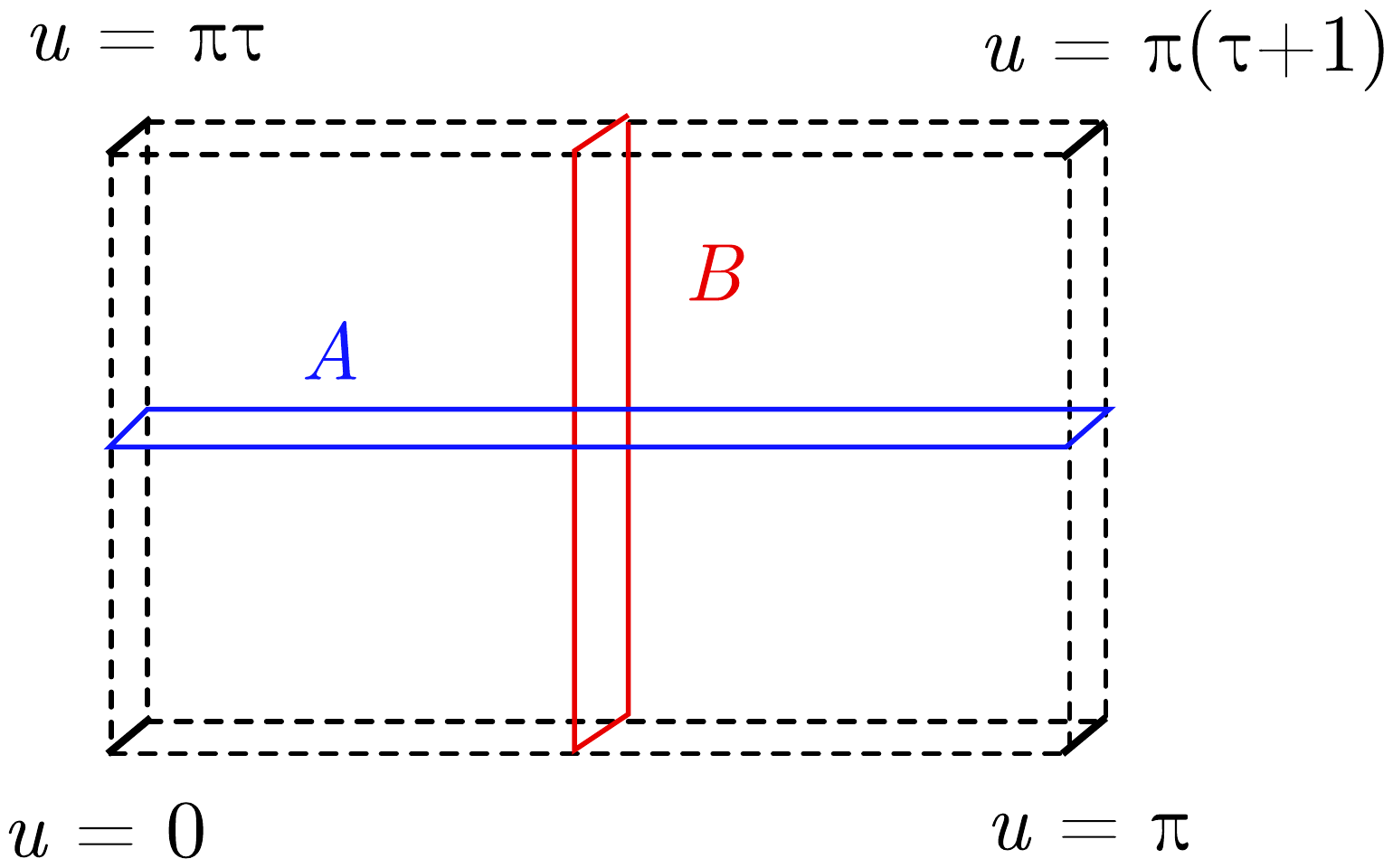}
    \caption{The $\mathbb{Z}_2$ quotient of the torus leads to a double-cover of the sphere which is flat except for conical defects at the corners, with uniformizing coordinates $u$ as indicated. The $A$ and $B$ cycles of the torus can be associated with bulk and boundary OPE channels in the (B)CFT, with insertions at the corners.}
    \label{fig:pillow}
\end{figure}

The parameter $\tau$ appearing in $q$ is the modulus of a torus which covers the Riemann sphere twice.
We will proceed with this construction, using Cardy's doubling trick to suppose we have a whole plane to play around with.
We will pick a torus $T^2$ which branches at $0, \eta, 1, \infty$, a Riemann surface described by the following equation:
\begin{equation}
    y^2 = x(x-\eta)(x-1),
\end{equation}
where $x, y \in \hat{\mathbb{C}}$ are points on the Riemann sphere.
This is the Weierstrass cubic associated with the lattice $\Lambda = \langle 1,\tau\rangle$ which quotients the complex plane to give the torus.
This provides a double cover of the sphere since the defining equation is invariant under $y \mapsto -y$, and the fixed points of this map are precisely the branch points.
The pillow has the topology of a sphere, and is flat, except for conical defects at these branch points, as depicted in Fig. \ref{fig:pillow}.



The fundamental domain of the torus is oblique for general $\tau$, but we can transform it into a rectangle using a \emph{uniformizing} coordinate $u$ defined by
\begin{equation}
    du = \frac{L}{\theta_3(q)^2}\frac{dx}{y}.
\end{equation}
This has width $2\pi L$, as one can check by performing the $x$ integral.
The $\mathbb{Z}_2$ action $y\mapsto -y$ becomes $u \mapsto -u$.
In the $u$ coordinates, the defects have coordinates
\begin{equation}
    u(x=0)=0, \quad u(x=\eta)=\pi, \quad u(x=1)=\pi(\tau+1), \quad u(x=\infty)= \pi\tau.
\end{equation}
We can cut the pillow in half in two ways: the horizontal $A$ cycle and the vertical $B$ cycle, which separate the corners into pairs, also shown in \ref{fig:pillow}.

\subsection{Evaluating the correlator}

We now consider how to implement this in the BCFT.
In the BCFT, $\eta = z_{12}z_{34}/z_{13}z_{24}$.
Thus, our insertions and their mirror images have the following identification on the pillow in $u$ coordinates:
\begin{equation}
    z_1 \mapsto 0, \quad z_{\bar{1}} \mapsto u = \pi, \quad z_{2} = \pi\tau, \quad z_{\bar{2}} = \pi(\tau+1).
\end{equation}
Thus, the boundary lies on the $B$ cycle, and we should quantize on the $A$ cycle.
If we normalize this cycle to have length $2\pi$ (or $\pi$ in the halved geometry), then the relevant Hamiltonian is just the dilatation operator in radial quantization (now on a half-cylinder), $H = L_0 - c/24$.

We evolve upwards by $\pi\tau$, i.e., with Euclidean time evolution operator 
\begin{equation}
    e^{i\pi\tau (L_0 - c/24)} = q^{L_0-c/24}.
\end{equation}
The change to $u$ coordinates is a Weyl rescaling, leading to an anomalous contribution to the correlator.
Performing this change and regularizing as in \cite{Maldacena:2015iua}, we obtain
\begin{align}
    \mathcal{F}(\eta) & = \Lambda(\eta) g(q) \\ g(q) & = \langle \calO(u=0)\calO(u=\pi\tau)\rangle_\text{pillow} \\
    \Lambda(\eta) & = \theta_3(q)^{c/2 -16h}[\eta(1-\eta)]^{c/24 - 2h}.
\end{align}

We can think of the pillow two-point function $g(q)$ as an expectation value
\begin{equation}
    g(q) = \langle \psi| q^{L_0-c/24}|\psi\rangle, \quad |\psi\rangle = \calO(0)|0\rangle,
\end{equation}
where due to our choice of quantization, $\langle\psi| = \langle 0| \calO(\pi\tau)$.
To be clear, here $|0\rangle$ is the vacuum state on the half-cylinder.
In the boundary channel, factoring out the $\Lambda$ prefactor also gives
\begin{equation}
    g(q) = \sum_{\hat{h}} |B_{\calO \hat{h}}|^2 \tilde{\mathcal{V}}_{\hat{h}}(q), \quad \tilde{\mathcal{V}}_{\hat{h}}(q) = \Lambda(\eta)^{-1}\mathcal{V}_{\hat{h}}(\eta).
\end{equation}
We can split the block into descendants:
\begin{equation}
    \tilde{\mathcal{V}}_{\hat{h}}(q) = \sum_{n\geq 0}a_n q^{n+\hat{h}-c/24}.
\end{equation}
In a unitary BCFT, the $a_n\geq 0$, since otherwise we can construct 
a linear combination of descendants with negative norm.

The bulk channel is naturally interpreted as quantizing on the $B$ cycle:
\begin{equation}
    g(q) = \langle \psi'|\bar{q}^{L_0 - c/24}|B\rangle,
\end{equation}
where $|\psi'\rangle = \calO(\pi\tau)\calO(0)|0\rangle$, $|0\rangle$ now the vacuum state for the full cylinder, and $\bar{q} = e^{-\pi i/\tau}$ is the $S$-transformed modular parameter.
Performing the bulk OPE expansion of our two operators, we end up with precisely the sum of bulk Virasoro primaries weighted by OPE coefficients and one-point functions given above.

\subsection{Seeking Lorentzian singularities}

\begin{figure}
    \centering
    \includegraphics[width=0.8\textwidth]{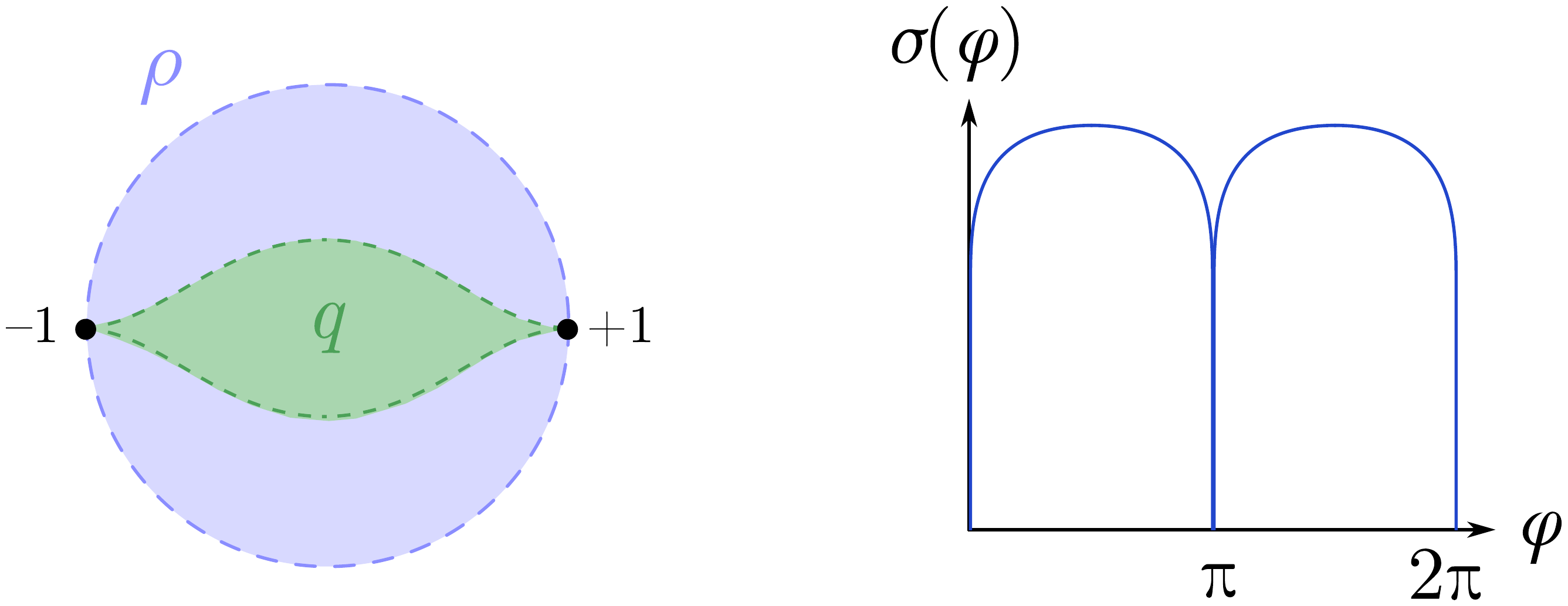}
    \caption{\emph{Left.} Relation between the radial $\rho$ variable and the nome $q$. \emph{Right.} The real part of $-\log q$ as a function of Lorentzian cross-ratio.}
    \label{fig:no-2d}
\end{figure}

Mapping the unit disk $|\rho| \leq 1$ to the $q$ variable leads a region sitting inside the unit disk of $q$, and hitting the boundary at $q = \pm 1$, which corresponds to $\rho = \pm 1$, and hence $\eta = 1, -\infty$, or $\xi = 0, -\infty$ in our preferred cross-ratio.
Note that $g(q)$ is finite inside the unit disk, since it is given by an expansion in powers of $q$ with positive, bounded
coefficients.
We depict this in Fig. \ref{fig:no-2d} (left) below.

We would like to use the behaviour of $g(q)$ to deduce that the only Lorentzian singularities are the light-cone singularities.  
Let us define
\begin{equation}
    \log q \left[\xi = -\cos^2 \left(\frac{\phi}{2}\right)\right] = -\sigma(\phi)+i\theta(\phi),
\end{equation}
where $\phi$ is the argument of $\rho$.
We plot the values of $\sigma(\theta) = -\log|q|$ in Fig. \ref{fig:no-2d} (right). It is positive except when $\phi \in \pi \mathbb{Z}$.
Recall that $\rho = r e^{i\phi} = e^{\tau + i\phi}$ in radial quantization, so that when we continue to the Lorentzian cylinder, $\tau = it$, the analytically continued nome becomes $q = e^{-\sigma(\phi + t)+i\theta(\phi + t)}$.
The Cauchy-Schwarz inequality then gives
\begin{equation}
    |g(q)| \leq \langle \psi | |q|^{L_0-c/24}|\psi\rangle = g(e^{-\sigma(\phi-t)}).
\end{equation}
This is the Euclidean pillow correlator again, which is finite except when $q = 1$, i.e., $\sigma = 0$, or $\phi + t \in \pi \mathbb{Z}$.
These are precisely the light-cone singularities.
We have thus proved that the only singularities in a 2D BCFT are the expected Euclidean and lightcone singularities, as advertised.



\section{Bulk Causality in Top-Down Model} \label{app:topdown}

In this appendix, we will endeavour to give further background information about causal features of the top-down construction mentioned in Section \ref{sec:top-down}, namely the AdS$_{2}$-cap solution of Chiodaroli, D'Hoker, and Gutperle. 
The metric functions for this solution are given by
\begin{equation}
    \begin{split}
        f_{1}^{4} & = \frac{2 \kappa^{2}}{\mu_{0}^{2}} \frac{\mu_{0}^{2} \Im(w)^{2} + \kappa^{2} | 1 - w^{2}|^{2}}{|w|^{4}} \\
        f_{2}^{4} & = \frac{2 \kappa^{2} \mu_{0}^{2} \Im(w)^{4}}{|w|^{4} \left( \mu_{0}^{2} \Im(w)^{2} + \kappa^{2} | 1 - w^{2}|^{2} \right)}\\
        \rho^{4} & = \frac{\mu_{0}^{2}}{8 \kappa^{2}} \frac{\mu_{0}^{2} \Im(w)^{2} + \kappa^{2} | 1 - w^{2}|^{2}}{|w|^{4} | 1 - w^{2}|^{4}} \: ,
    \end{split}
\end{equation}
where $\kappa > 0$ and $\mu$ are two real parameters. We will sometimes denote $a \equiv \frac{2 \kappa}{\mu_{0}}$ for convenience. The asymptotically AdS$_{3} \times S^{3}$ region is located at $w = 0$, whereas the AdS-cap singularities are located at $w = \pm 1$.

Let $(u, t)$ be the Poincar{\'e} coordinates of the AdS$_{2}$ factor and $(\theta, \phi)$ be the angular coordinates of the $S^{2}$ factor, so that
\begin{equation}
    ds_{\textnormal{AdS}_{2}}^{2} = \frac{du^{2} - dt^{2}}{u^{2}} \: , \qquad ds_{S^{2}}^{2} = d \theta^{2} + \sin^{2} \theta d \phi^{2} \: .
\end{equation}
It is straightforward to demonstrate that, in the limit that we approach the asymptotic boundary, the AdS$_{2}$ coordinates $(u, t)$ agree with the BCFT coordinates $(x_{\perp}, \hat{t})$ (arising in Fefferman-Graham gauge) up to a constant factor
\begin{equation}
    u \big|_{\partial} = - x_{\perp} / a \: , \qquad t \big|_{\partial} = \hat{t} / a \: .
\end{equation}
In particular, these factors will drop out of a calculation of the BCFT cross-ratio, so we can calculate the cross-ratio directly using the $(u, t)$ parameters.

We would like to shoot in geodesics from the asymptotic boundary of the AdS$_{3} \times S^{3}$ region, provided initial data 
consistent with the null constraint
\begin{equation}
    0 = \frac{f_{1}^{2}}{u^{2}} \left( \dot{u}^{2} - \dot{t}^{2} \right) + f_{2}^{2} \left( \dot{\theta}^{2} + \sin^{2} \theta \dot{\phi}^{2} \right) + \rho^{2} \left( \dot{x}^{2} + \dot{y}^{2} \right) \: ,
\end{equation}
and solve the geodesic equation to determine the intersection of the bulk light cone with the asymptotic boundary.

Before proceeding to determine the null geodesics, we may choose to multiply the metric functions by some fixed conformal factor, since this will not affect the bulk light cone structure; we therefore choose for convenience to multiply the metric by $f_{1}^{-2}$, to obtain
\begin{equation}
    d\tilde{s}^{2} = ds_{\textnormal{AdS}_{2}}^{2} + \frac{f_{2}^{2}}{f_{1}^{2}} ds_{S^{2}}^{2} + \frac{\rho^{2}}{f_{1}^{2}} dw d\bar{w} \: .
\end{equation}
Our choice is motivated by the fact that $\rho^{2} / f_{1}^{2}$ is a simple function of the complex coordinates $(w, \bar{w})$, namely
\begin{equation}
    \frac{\rho^{2}}{f_{1}^{2}} = \frac{1}{a^{2}} \frac{1}{|1 - w^{2}|^{2}} \: .
\end{equation}
With this choice, and taking fixed $\theta = \frac{\pi}{2}$ without loss of generality, we find the components of the geodesic equation decouple pairwise, giving
\begin{equation} \label{eq:geodesic_1}
    \begin{split}
        \ddot{x} & = \frac{4 y (x^{2} + y^{2} + 1)}{\left( (x-1)^{2} + y^{2} \right) \left( (x+1)^{2} + y^{2} \right)} \dot{x} \dot{y} + \frac{2 x (x^{2} + y^{2} - 1)}{\left( (x-1)^{2} + y^{2} \right) \left( (x+1)^{2} + y^{2} \right)} \left( \dot{x}^{2} - \dot{y}^{2} \right) \\
        & \qquad - 8 x y^{2} \dot{\phi}^{2} \left( \frac{\left( x^{2} + y^{2} - 1 \right) \left( \left( x + 1 \right)^{2} + y^{2} \right) \left( \left( x - 1 \right)^{2} + y^{2} \right)  }{\left( (1 - x^{2})^{2} + y^{4} + 2 \left( x^{2} + 1 + \frac{\mu_{0}^{2}}{2 \kappa^{2}} \right) y^{2} \right)^{2}} \right) \: , \\
        \ddot{y} & = \frac{4 x (x^{2} + y^{2} - 1)}{\left( (x-1)^{2} + y^{2} \right) \left( (x+1)^{2} + y^{2} \right)} \dot{x} \dot{y} - \frac{2 y (x^{2} + y^{2} + 1)}{\left( (x-1)^{2} + y^{2} \right) \left( (x+1)^{2} + y^{2} \right)} \left( \dot{x}^{2} - \dot{y}^{2} \right) \\
        & \qquad + 4 y \dot{\phi}^{2} \left( \frac{\left( \left( x^{2} - 1 \right)^{2} - y^{4} \right) \left( \left( x + 1 \right)^{2} + y^{2} \right) \left( \left( x - 1 \right)^{2} + y^{2} \right)  }{\left( (1 - x^{2})^{2} + y^{4} + 2 \left( x^{2} + 1 + \frac{\mu_{0}^{2}}{2 \kappa^{2}} \right) y^{2} \right)^{2}} \right) \\
        \frac{\ddot{\phi}}{\dot{\phi}} & = \left( \frac{4 x^{2} \left( x^{2} + y^{2} - 1 \right)}{ (1 - x^{2})^{2} + y^{4} + 2 \left( x^{2} + 1 + \frac{\mu_{0}^{2}}{2 \kappa^{2}} \right) y^{2}} \right) \frac{\dot{x}}{x} \\
        & \qquad - \left( \frac{2  \left( \left( x^{2} - 1 \right)^{2} - y^{4} \right) }{(1 - x^{2})^{2} + y^{4} + 2 \left( x^{2} + 1 + \frac{\mu_{0}^{2}}{2 \kappa^{2}} \right) y^{2}} \right) \frac{\dot{y}}{y}  \: ,
    \end{split}
\end{equation}
and 
\begin{equation} \label{eq:geodesic_2}
    \ddot{u} = \frac{1}{u} \left( \dot{u}^{2} + \dot{t}^{2} \right) \: , \quad \ddot{t} = \frac{2}{u} \dot{u} \dot{t} \: ,
\end{equation}
while the null condition is still as above (with $\dot{\theta}=0$).

\subsection{Short Light-Crossing Time with Fine-Tuned Initial Conditions}

Before considering more general initial data, we will consider the case $\dot{x}(0) = 0$; it is straightforward to see that $x(s)=0$ for the entire trajectory in this case, so that the equations simplify to
\begin{equation}
    \begin{split}
        \frac{\ddot{y}}{y} & = \frac{2 \dot{y}^{2}}{\left( 1 + y^{2} \right) } + 4 \dot{\phi}^{2} \left( \frac{\left( 1 - y^{4} \right) \left( 1 + y^{2} \right)^{2}  }{\left( 1 + \left( 1 + \frac{\mu_{0}^{2}}{2 \kappa^{2}} \right) y^{2} + y^{4}  \right)^{2}} \right) \\
        \frac{\ddot{\phi}}{\dot{\phi}} & = - \left( \frac{2  \left(1 - y^{4} \right) }{ 1 + \left( 1 + \frac{\mu_{0}^{2}}{2 \kappa^{2}} \right) y^{2} + y^{4}  } \right) \frac{\dot{y}}{y}  \: .
    \end{split}
\end{equation}
We can integrate out the $\phi$ equation, which gives
\begin{equation}
\begin{split}
    \ln ( \dot{\phi}(s) / \dot{\phi}(s_{0}))  & = - \int_{s_{0}}^{s} ds' \: \left( \frac{2  \left(1 - y(s')^{4} \right) }{ 1 + \left( 1 + \frac{\mu_{0}^{2}}{2 \kappa^{2}} \right) y(s')^{2} + y(s')^{4}  } \right) \frac{\dot{y}(s')}{y(s')} \\
    & = \Bigg[ \ln \left( \frac{y(s)^{4} + \left( 1 + \frac{\mu_{0}^{2}}{2 \kappa^{2}} \right) y(s)^{2} + 1}{y(s)^{2}} \right) \Bigg]^{s}_{s_{0}} \: ,
\end{split}
\end{equation}
that is,
\begin{equation}
    \dot{\phi}(s) =  c \cdot  \frac{y(s)^{4} + \left( 1 + \frac{\mu_{0}^{2}}{2 \kappa^{2}} \right) y(s)^{2} + 1}{y(s)^{2}} \: , \qquad c \equiv \frac{ \dot{\phi}_{0} y_{0}^{2}}{y_{0}^{4} + \left( 1 + \frac{\mu_{0}^{2}}{2 \kappa^{2}} \right) y_{0}^{2} + 1} \: ,
\end{equation}
where the subscript zero denotes evaluation at a reference value $s = s_{0}$. 
The $y$ equation then gives
\begin{equation} \label{eq:y_finetune}
    \begin{split}
        & \frac{\ddot{y}}{y} - \frac{2 \dot{y}^{2}}{\left( 1 + y^{2} \right) } = \frac{4 c^{2}}{y^{4}} (1 - y^{4})(1 + y^{2})^{2}  \: ,
    \end{split}
\end{equation}
where $y(s_{0}) = y_{0}$ and $\dot{\phi}(s_{0}) = \dot{\phi}_{0}$. Note that the parameters $\mu_{0}, \kappa$ have dropped out of this expression.


For any choice of initial conditions for this ODE, the trajectory diverges to $y = \infty$ in some finite affine parameter. The point $\infty$ on $\Sigma$ is a regular point of the smooth 6D geometry; in particular, making coordinate transformation
\begin{equation}
    v = \frac{1}{w} \: , \qquad \bar{v} = \frac{1}{\bar{w}} \: ,
\end{equation}
and letting
\begin{equation}
    v = r e^{i \varphi} \: , \qquad \bar{v} = r e^{-i \varphi} \: ,
\end{equation}
we have for small $r$
\begin{equation}
    \begin{split}
        ds^{2} & \approx \frac{\mu_{0}}{\sqrt{8}} \left( \frac{dr^{2}}{r^{2}} + \frac{a^{2}}{u^{2}} \left( du^{2} - dt^{2} \right) \right) + \frac{\mu_{0}}{\sqrt{8}} \left( d \varphi^{2} + 4 r^{2} \sin^{2} \varphi \left( d \theta^{2} + \sin^{2} \theta d\phi^{2} \right) \right) \: .
    \end{split}
\end{equation}
Evidently, this point is at an infinite proper distance from any other point in the geometry that we might choose, but is reached by our null geodesic in finite affine parameter; the metric in the vicinity of this point is conformal to 
\begin{equation}
    \begin{split}
        ds^{2} & = \left( dr^{2} + r^{2} d \varphi^{2} \right) + O(r^{2}) \: ,
    \end{split}
\end{equation}
so our geodesic will ``turn around" at this point.

For concreteness, we can consider the case $\dot{\phi}(0) = 0$; the corresponding geodesic leaves and returns to the asymptotic boundary $y=0$ in affine parameter $s = \frac{\pi}{\dot{y}(0)}$.
Indeed, the ODE determining the trajectory on the Riemann surface $\Sigma$
\begin{equation}
    \ddot{y} = \frac{2 y}{(y^{2} + 1)} \dot{y}^{2} \: , \qquad y(0) = 0 \: , \: \dot{y}(0) = y_{1} \: ,
\end{equation}
has solution
\begin{equation}
    y(s)  = \begin{cases}
        \tan(y_{1} s) & 0 < s < \frac{\pi}{2 y_{1}} \\
        \tan\left( \pi - y_{1} s \right) & \frac{\pi}{2 y_{1}} < s < \frac{\pi}{y_{1}} 
    \end{cases} \: .
\end{equation}
In particular, we see that the geodesic returns to the asymptotic boundary in affine parameter $s = \frac{\pi}{y_{1}}$. 
Without loss of generality, we take $y_{1} = 1$ in the following, which is simply a choice of normalization for the affine parameter. 

We can now turn to the coordinates $(u, t)$. The general solution to the differential equations for these parameters is
\begin{equation}
    \begin{split}
        u(s) & = u(s_{0}) \sec \left( \frac{(s - s_{0})}{a}  \right) \\
        t(s) & = u(s_{0}) \tan \left( \frac{(s - s_{0})}{a}  \right) + u(s_{0}) \tan \left( \frac{s_{0}}{a} \right) \: .
    \end{split}
\end{equation}
The only way that this trajectory couples to the trajectory on $\Sigma$ is that the latter determines the total affine parameter $S$ elapsed along the trajectory. 
Thus, in particular, we may calculate the cross-ratio for the initial and final coordinates on the geodesic, finding
\begin{equation}
    \begin{split}
        \xi & = \frac{-\left( t(S) - t(0) \right)^{2} + \left( u(S) - u(0) \right)^{2}}{4 u(0) u(S)} = \frac{1}{2} \left( \cos \left( S / a \right) - 1 \right) \: .
    \end{split}
\end{equation}
In terms of the ``radial coordinate" cross-ratio $\rho$ defined by
\begin{equation}
    \xi = \frac{(1 - \rho)^{2}}{4 \rho} \: ,
\end{equation}
we have
\begin{equation}
    \rho = \cos \left( S / a \right) \pm i \sin \left( S / a \right) = e^{\pm \frac{i S}{a}} \: ,
\end{equation}
where for example $S = \pi$ for the case $\dot{\phi}(0) = 0$. 
Consequently, if we expect this cross-ratio to describe the locus of a Lorentzian singularity of a BCFT two-point function, then we might expect that such singularity can be attributed to the contribution of a particular tower of boundary operators with dimensions exhibiting an asymptotic spacing which is an integer multiple of $2a$. 

\clearpage

\subsection{Geodesics for General Initial Conditions}

\begin{figure}
    \centering
    \includegraphics[height=6.5cm]{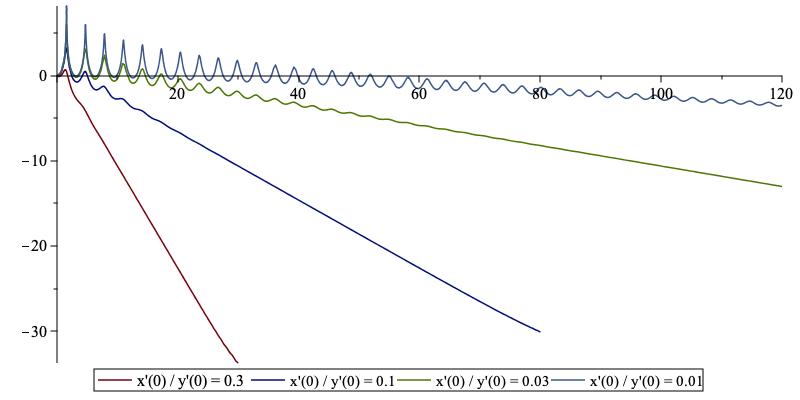}
    \caption{Logarithmic distance to the AdS-cap $\ln\left( (x(s) - 1)^{2} + y(s)^{2} \right)$ versus affine parameter $s$ for geodesics with various initial conditions on the internal space $\Sigma$. Here we have chosen $\mu_{0} = \kappa = 1$ for concreteness. }
    \label{fig:cap_approach}
\end{figure}

\begin{figure}
    \centering
    \includegraphics[height=6.5cm]{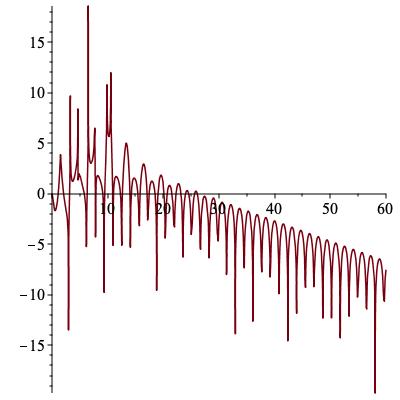} \qquad
    \includegraphics[height=6.5cm]{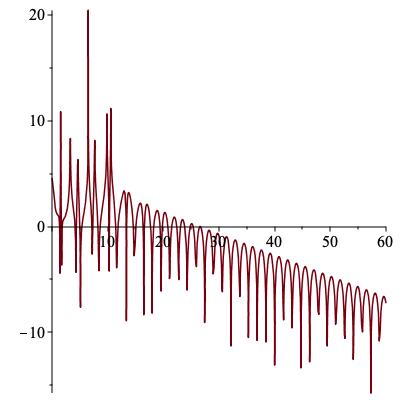}
    \caption{Plots of $\ln \left( \frac{\dot{\epsilon}_{x}^{2} / \epsilon}{\dot{\epsilon}^{2} / \epsilon^{2}} \right)$ (left) and $\ln \left( \frac{\dot{\epsilon}_{y}^{2} / \epsilon}{\dot{\epsilon}^{2} / \epsilon^{2}} \right)$ (right) versus affine parameter for a geodesic with initial condition $\dot{x}(0) / \dot{y}(0) = 0.3$; these quantities are required to be small for ingoing geodesics to approach the cap exponentially in the affine parameter. Here we have chosen $\mu_{0} = \kappa = 1$ for concreteness. }
    \label{fig:cap_vicinity}
\end{figure}

More general solutions to the geodesic equations (\ref{eq:geodesic_1}),  (\ref{eq:geodesic_2}) and the null constraint, without the assumption $x(s) = 0$, may also be studied numerically. Doing so, we appear to find that null geodesics with initial momentum in the $x$-direction quite generically approach the AdS-caps. In Figure \ref{fig:cap_approach}, we plot the distance of various geodesics from the curvature singularity as a function of affine parameter for various initial conditions, restricting momentarily to the case with no momentum on the $S^{2}$. This is consistent with the conclusion of our perturbative argument in the main text; indeed, the perturbative assumption made there appears to hold in general, with an example of this fact illustrated in Figure \ref{fig:cap_vicinity}. 

\begin{figure}
    \centering
    \includegraphics[height=6.5cm]{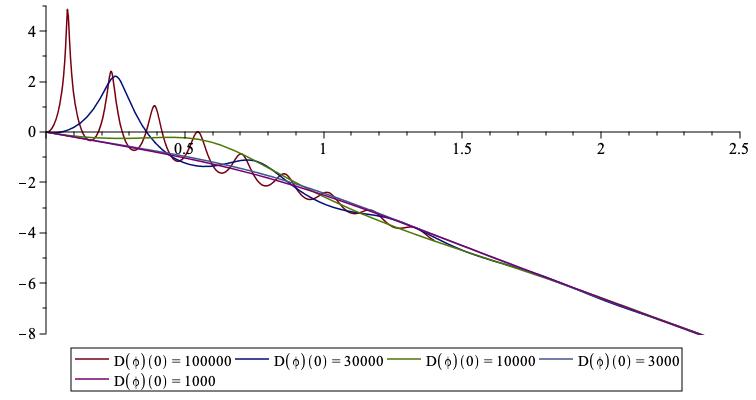}
    \caption{Logarithmic distance to the AdS-cap $\ln\left( (x(s) - 1)^{2} + y(s)^{2} \right)$ versus affine parameter $s$ for geodesics with various initial conditions on the $S^{2}$. We take $x = y = 10^{-4}$ as the origin for these geodesics, with initial condition $\dot{x}(0) = \dot{y}(0) = 1$. Here we have chosen $\mu_{0} = \kappa = 1$ for concreteness.}
    \label{fig:various_phi1}
\end{figure}

We can attempt to extend this analysis to the case with momentum on the $S^{2}$.
However, we cannot actually send in geodesics from the asymptotic boundary with momentum on the $S^{2}$; the system of ODEs is singular at $x = y = 0$, and the angular momentum on $S^{2}$ diverges there in general. 
Instead, we choose a point near the the origin of $\Sigma$ and send a geodesic in from this point with initial momentum on the $S^{2}$, with the hopes that this will capture the relevant behaviour. Doing so, we again find that the AdS-caps act as attractors; see Figure \ref{fig:various_phi1} for an example.

\newpage

\providecommand{\href}[2]{#2}\begingroup\raggedright\endgroup

\end{document}